\documentclass{article}
\usepackage{arxiv}
\usepackage{graphicx,amsmath,amsfonts,amscd,amsthm,amssymb,pstricks}
\usepackage{graphicx,color}
\usepackage{dcolumn}
\usepackage{bm}
\usepackage{longtable}
\usepackage{amsmath}
\usepackage{mathrsfs}
\usepackage{amsfonts}
\usepackage{textcomp}
\usepackage{esvect}
\usepackage{float}
\usepackage[USenglish,british]{babel}
\usepackage{environ}
\usepackage{xifthen}
\usepackage{xargs}			
\usepackage{hyperref}
\usepackage{physics,mathtools}
\usepackage[separate-uncertainty = true]{siunitx}
\usepackage{multirow}
\usepackage{comment}

\newcommand{\av}[1]{\left<#1\right>}
\newcommand{\ie}{i.e.\ }

\newcommand{\eg}{e.g.\ }

\newcommand{\wrt}{w.r.t.\ }

\newcommand{\phia}{\phi_\mathrm{a}}
\newcommand{\phib}{\phi_\mathrm{b}}
\newcommand{\Ja}{J_\mathrm{a}}
\newcommand{\Jb}{J_\mathrm{b}}

\def\block(#1,#2)#3{\multicolumn{#2}{c}{\multirow{#1}{*}{$ #3 $}}}
\allowdisplaybreaks[1]

\title{On the adiabaticity of emittance exchange due to crossing of the coupling resonance}

\author{A. Bazzani \\
Dipartimento di Fisica e Astronomia, Universit\`a di Bologna, via Irnerio 46, Bologna, Italy \\
\And
F. Capoani \\
Dipartimento di Fisica e Astronomia, Universit\`a di Bologna, via Irnerio 46, Bologna, Italy \\
and\\
Beams Department, CERN, 1211 Geneva 23, Switzerland \\
\And
M. Giovannozzi\thanks{Corresponding author: massimo.giovannozzi@cern.ch} \\
Beams Department, CERN, 1211 Geneva 23, Switzerland \\
\And
A.I. Neishtadt \\
Department of Mathematical Sciences, Loughborough University, Loughborough LE11 3TU, UK \\
and\\
Space Research Institute of RAS, Moscow 117997, Russia}

\begin{document}
\maketitle

\begin{abstract}
In circular accelerators, crossing the linear coupling  resonance induces the exchange of the transverse emittances, provided the process is adiabatic. This has been considered in some previous works, where the description of the phenomenon has been laid down, and, more recently, where a possible explanation of the numerical results has been proposed. In this paper, we introduce a theoretical framework to analyze the crossing process, based on the theory of adiabatic invariance for Hamiltonian mechanics, which explains in detail various features of the emittance exchange process.
\end{abstract}

\section{Introduction}
The impact of linear coupling on transverse betatron motion has been extensively studied, as it has a peculiar impact already on the linear dynamics. In 2001, the phenomenon of emittance exchange due to dynamic crossing of the difference coupling resonance was studied~\cite{Metral:529690}, with further results reported in 2007~\cite{PhysRevSTAB.10.064003}, in which it is mentioned that the full emittance exchange happens if the resonance crossing is adiabatic and an adiabatic condition is given. This research has opened a new domain of investigations and a recent paper addressed the same topic with the goal to develop a complete theory to describe the emittance exchange process~\cite{PhysRevAccelBeams.23.044003}. 

In recent years, there has been intense theoretical efforts to study in detail the phenomenon of resonance crossing in one degree-of-freedom (1DoF) Hamiltonian systems in view of devising novel beam manipulations~\cite{PhysRevLett.88.104801,PhysRevSTAB.7.024001,PhysRevSTAB.10.034001,PhysRevSTAB.12.024003,PhysRevE.89.042915,PhysRevAccelBeams.20.121001}. This culminated with the proposal and final implementation of the CERN PS Multi-Turn Extraction (MTE) as an operational means to provide an optimized extraction technique based on nonlinear beam dynamics~\cite{PhysRevSTAB.9.104001,PhysRevSTAB.12.014001,Borburgh:2137954,PhysRevAccelBeams.20.014001,PhysRevAccelBeams.20.061001,PhysRevAccelBeams.22.104002,Vadai:2702852}. 

A natural extension of what was done with MTE is the analysis of a 2DoF nonlinear system that crosses a 2D nonlinear resonance. Inspired by Ref.~\cite{PhysRevLett.110.094801}, some very promising results have been obtained~\cite{2Dmanipulations}, which indicate that the adiabatic crossing of resonances can be an efficient means to manipulate the invariants of a Hamiltonian system. This effect gives the possibility of redistributing the transverse emittances between the transverse degrees of freedom. It is worth stressing that the mathematical framework for these studies is the  theory of adiabatic invariance for Hamiltonian systems\cite{NEISHTADT198158,Arnold:937549}.

This framework provides also the natural way of addressing the analysis of the resonance crossing in the presence of linear coupling. In this paper, we show how all observations reported in previous works, such as~\cite{PhysRevSTAB.10.064003, PhysRevAccelBeams.23.044003}, find a clear explanation using the results of adiabatic theory. Furthermore, we extend the analysis to the case in which nonlinear detuning with amplitude is present in the considered system.

The plan of the paper is the following: in Section~\ref{sec:hammod}, the coupling Hamiltonian model is introduced and discussed in detail (Section~\ref{sec:skew}), including an original view of the phase space on a sphere (Section~\ref{sec:sphere}). In Section~\ref{sec:digression}, the same Hamiltonian system is analyzed using the normal modes and the main results on the properties of the dynamics are derived, whereas in Section~\ref{sec:detuning}, the analysis of the effect of detuning with amplitude on the original Hamiltonian is carried out. A digression is made in Section~\ref{sec:digr}, where the problem of two-way crossing of the coupling resonance is considered. In Section~\ref{sec:map}, the map model is introduced and, in Section~\ref{sec:res}, the results of the numerical simulations are presented and discussed in detail. Finally, conclusions are drawn in Section~\ref{sec:conc}, while some mathematical details are reported in the Appendices~\ref{sec:app1},~\ref{sec:app2},~\ref{sec:app3}, and~\ref{sec:app4}.
\section{The Hamiltonian model and its dynamics}\label{sec:hammod}
Following the treatment used in  Refs.~\cite{PhysRevE.49.2347,Lee:2651939,PhysRevLett.110.094801}, we consider a Hamiltonian written in the following form
\begin{equation}
  H(p_x,p_y,x,y) = \frac{p_x^2+p_y^2}{2} + \frac{1}{2} \left ( \omega_x^2\,  x^2+\omega_y^2 \, y^2+2 q \, x \, y \right ) \, ,
  \label{eq:hamq}
\end{equation}
where $q = -\sqrt{\beta_x\beta_y} \, \hat{q}$, and the coefficient $\hat q$ is defined as 
\begin{equation}
\hat{q}= \frac{1}{2 B \rho} \left ( \frac{\partial B_y}{\partial y}+\frac{\partial B_x}{\partial x} \right )
\end{equation}
and represents the effect of a skew quadrupole on the betatron dynamics. In the following, the notation $z$ will be used to denote either the coordinate $x$ or $y$.
\subsection{Analysis of the dynamics in the presence of a skew quadrupole}\label{sec:skew}
We consider the adiabatic crossing of the linear coupling resonance, namely $\omega_x-\omega_y=0$, when the frequencies are slowly modulated, and we define 
\begin{equation}
    \delta(\lambda)=\omega_x(\lambda)-\omega_y(\lambda)
\end{equation}
with $\lambda=\epsilon t$, $\epsilon \ll 1$, and $\epsilon$ is the adiabatic parameter that describes the resonance crossing process. Without loss of generality, $\delta(\lambda)$ is defined by a linear function that varies from positive to negative values (or vice versa) crossing zero. 

The eigenvalues of the matrix associated to the quadratic potential matrix are given by 
\begin{equation}
\omega^2_{1,2} =\frac{\omega_x^2+\omega_y^2\pm \sqrt{(\omega_x^2-\omega_y^2)^2+4 q^2}}{2}  
\label{eq:normmod}
\end{equation}
and it is convenient to define
\begin{equation}
    \delta_2(\lambda)=\omega_x^2(\lambda)-\omega_y^2(\lambda)=\delta(\lambda) \left ( \omega_x(\lambda)+\omega_y(\lambda) \right )
\end{equation}
so that
\begin{eqnarray}
\omega_1^2(\lambda)&=&\omega_x^2(\lambda)-\frac{\delta_2(\lambda)- \sqrt{\delta_2^2(\lambda)+4q^2}}{2} \nonumber \\
& & \\
\omega_2^2(\lambda)&=&\omega_y^2(\lambda)+\frac{\delta_2(\lambda)-\sqrt{\delta_2^2(\lambda)+4q^2}}{2} \nonumber \, .
\end{eqnarray}

The corresponding eigenvectors are
\begin{eqnarray}
\bm v_1(\lambda)&=&c_1 \left ( \frac{\delta_2(\lambda)+\sqrt{\delta_2^2(\lambda)+4q^2}}{2}, q\right ) \nonumber \\
& & \\
\bm v_2(\lambda)&=&c_2 \left ( -q, \frac{\delta_2(\lambda)+\sqrt{\delta_2^2(\lambda)+4q^2}}{2}\right ) \nonumber \, ,
\end{eqnarray}
where $c_i$ are the normalising constants. Note that for $q\ll 1$ and $\delta_2(\lambda)>0$ one has $\bm v_1\to \bm e_x$ and $\bm v_2\to \bm e_y$, where $\bm e_x, \bm e_y$ are the unit vectors defining the horizontal and vertical planes. When $\delta_2(\lambda)=0$, \ie $\omega_x(\lambda)=\omega_y(\lambda)$, then $\bm v_1$ and $\bm v_2$ define the two bisectors of the two angles defined by the horizontal axis and the positive vertical axis, whereas when $|q|\ll 1$ and $\delta_2(\lambda)<0$, then $\bm v_1\to \bm e_y$ and $\bm v_2\to -\bm e_x$. Therefore, the passage through the resonance $\omega_x-\omega_y$ implies an exchange of the direction of the eigenvectors. 

The value of $q$ is constrained by the conditions that $\omega_{1,2}$ are both real, \ie
\begin{equation}
|q| \leq \omega_x \, \omega_y
\end{equation}
otherwise the closed orbit, corresponding to the fixed point at the origin, becomes unstable.

It is also worth noting that the following relations hold
\begin{equation}
\begin{split}
\omega^2_1+\omega^2_2 & = \omega^2_x+\omega^2_y \\
\omega^2_1-\omega^2_2 & = \sqrt{\left (\omega^2_x-\omega^2_y \right )^2 +4 q^2} \\
\omega_1\omega_2& =\sqrt{\omega_x^2\omega_y^2-q^2}
\label{eq:tune_rel}
\end{split}
\end{equation}
from which one remarks that the difference resonance cannot be crossed by $\omega_{1,2}$ as the eigenvalues cannot get closer than $(\omega^2_1-\omega^2_2)_{\rm min} = 2 |q|$ as it is well-known (see, \eg Refs.~\cite{PhysRevSTAB.10.064003,PhysRevAccelBeams.23.044003} and references therein). This observation leads to an essential conclusion: in the physical co-ordinates, the coupling resonance can be crossed, but the tunes are not the eigenvalues of the system. On the other hand, in the co-ordinate system of the eigenvalues the resonance cannot be crossed, although the eigenvalues are the proper quantities to describe the dynamics. For this reason, the term pseudo-resonance crossing will be also used in the following.

We introduce the linear normal form for the Hamiltonian~\eqref{eq:hamq} and the dependence of the symplectic transformation on time (via the parameter $\lambda$) introduces a further term in the original Hamiltonian. If we indicate with $\bm{G}(\lambda)$ the matrix of the transformation $Z =z \, \sqrt{\omega_z(\lambda)}$, it induces the  transformation  
\begin{equation}
\bm x=\bm G(\lambda) \, \bm X\, ,
\end{equation}
where $\bm X$ are the new co-ordinates. A generating function $F_2( \bm x,\bm P, \lambda)$ for the symplectic transformation can be written in the form
\begin{equation}
F_2(\bm x,\bm P, \lambda)=\bm P \, \bm G(\lambda)^{-1} \bm x 
\end{equation}
and the new Hamiltonian reads
\begin{equation}
\begin{split}
H(\bm X,\bm P,\lambda) &=\omega_x(\lambda)\frac{X^2+ P_x^2}{2}+\omega_y(\lambda)\frac{Y^2+P_y^2}{2}+\\
& +\frac{q}{\sqrt{\omega_x(\lambda) \omega_y(\lambda)}} X\, Y +\epsilon \, \bm P \pdv{\, \bm{G}^{-1}}{\lambda} \bm G \bm X \, ,
\end{split}
\end{equation}
where the last term is the time derivative of the generating function. The final form of the Hamiltonian reads
\begin{equation}
    \begin{split}
H(\bm X, \bm P, \lambda) & =\omega_x(\lambda)\frac{X^2+P_x^2}{2}+\omega_y(\lambda) \frac{Y^2+P_y^2}{2}+\\
& + \frac{q}{\sqrt{\omega_x(\lambda)\omega_y(\lambda)}} X \, Y + \\
& + \frac{\epsilon}{2}\left [\frac{\omega_x'(\lambda)}{\omega_x(\lambda)}X\, P_x + \frac{\omega_y'(\lambda)}{\omega_y(\lambda)}Y \, P_y\right ] \, ,
    \end{split}
    \label{eq:hamt}
\end{equation}
where $\omega'=d\omega/d\lambda$. The linear action-angle variables $(\bm \theta, \, \bm I)$ can be used to recast the Hamiltonian~\eqref{eq:hamt} in the form
\begin{equation}
    \begin{split}
H(\bm \theta,\bm I, \lambda)&= \omega_x(\lambda) I_x+\omega_y(\lambda) I_y+
\frac{2q}{\sqrt{\omega_x(\lambda)\omega_y(\lambda)}} \times \\
& \times \sqrt{I_x I_y}\sin\theta_x\sin\theta_y + \epsilon\left [\frac{\omega_x'(\lambda)}{\omega_x(\lambda)}I_x\sin\theta_x\cos\theta_x+ \right . \\
& + \left . \frac{\omega_y'(\lambda)}{\omega_y(\lambda)}I_y\sin\theta_y\cos\theta_y\right ]
    \end{split} \, .
\end{equation}

We remark that the Hamiltonian dynamics is singular~\footnote{Here and in the following, \textit{singular dynamics} means that the corresponding equations of motion are singular.} at $I_x=I_y=0$ and the frequencies $\omega_{x,y}$ are not the linear frequencies around the elliptic fixed point due to the presence of the linear coupling term. Hence, the condition $\omega_x(\lambda)=\omega_y(\lambda)$ is not a true dynamical resonance condition. 

The Hamiltonian contains two small parameters, namely $\epsilon \to 0$ in the adiabatic limit and $q$ that measures the strength of the linear coupling: the main issue is how to determine the interplay between the two small parameters in the limit $\epsilon\to 0$. 

The introduction of a slow phase $\phia=\theta_x-\theta_y$ using the generating function
\begin{equation}
    F_2(\bm \theta, \bm J)=\begin{pmatrix}
\Ja, & \Jb
\end{pmatrix}
\begin{pmatrix}
1 &          - 1 \\
0 & \phantom{-}1
\end{pmatrix}
\begin{pmatrix}
\theta_x\\ \theta_y
\end{pmatrix}
\end{equation}
transforms the Hamiltonian to the form
\begin{equation}
    \begin{split}
        H(\bm \phi, \bm J,\lambda) & =\delta(\lambda) \Ja+\omega_y \Jb+
\frac{2q}{\sqrt{\omega_x(\lambda)\omega_y(\lambda)}} \times \\
& \times \sqrt{\Ja (\Jb-\Ja)}\sin(\phia+\phib)\sin\phib+ \\
&+
\epsilon\left [\frac{\omega_x'(\lambda)}{\omega_x(\lambda)}\Ja\sin(\phia+\phib)\cos(\phia+\phib)+ \right . \\
& + \left . \frac{\omega_y'(\lambda)}{\omega_y(\lambda)}(\Jb-\Ja)\sin\phib\cos\phib\right ]
    \end{split}
\end{equation}
and since we focus on the analysis when $\delta(\lambda)\to 0$, it is possible to apply a perturbative approach averaging over the fast-evolving angle $\phib$ to obtain the Hamiltonian
\begin{equation}
    \begin{split}
        H(\bm \phi, \bm J,\lambda) & =\delta(\lambda) \Ja+\omega_y \Jb+
\frac{q}{\sqrt{\omega_x(\lambda)\omega_y(\lambda)}} \times \\
& \times \sqrt{\Ja (\Jb-\Ja)}\cos\phia+
O(\epsilon^2)+O(q^2)
    \end{split} \, .
\end{equation}

Without any further assumption, it follows that $\Jb$ is invariant up to an error $O(q^2) + O(\epsilon^2)$ for a time interval of order $O(\epsilon^{-1})$. The perturbative  approach is possible only if this error is small, so that $\Jb$ can be considered constant during the resonance-crossing process. In such a case, the action of the 1DoF Hamiltonian
\begin{equation}
\begin{split}
    H(\bm \phi,\bm J,\lambda) & = \delta(\lambda) \Ja+\frac{q}{\sqrt{\omega_x(\lambda)\omega_y(\lambda)}} \times \\
    & \times \sqrt{\Ja (\Jb-\Ja)}\cos\phia
\end{split}
\label{hamsphe}
\end{equation}
can be considered an adiabatic invariant up to an error $O(q^2\,\epsilon^{-1})$ for a time interval $O(\epsilon^{-1})$, and we can study the change of $\Ja$ when $\delta(\lambda)$ passes through zero. In the end, it is possible to restrict the problem of studying the resonance-crossing process for the original Hamiltonian~\eqref{eq:hamq} by considering the dynamics generated by $H$ that can be recast in the following form
\begin{equation}
H(\phi, J,\lambda)=\delta(\lambda)J+ q \sqrt{(1-J)J}\sin\phi \, ,
\label{eq:ham_fundamental}
\end{equation}
where, without loss of generality, we have re-scaled the action according to $J=\Ja/\Jb$ so that $J=0$ and $J=1$ are singular lines for the Hamiltonian~\eqref{eq:ham_fundamental}, we also defined $\phi=\phia+\pi/2$, and then replaced $\delta(\lambda) \to \delta(\lambda) \, \sqrt{\omega_x(\lambda)\omega_y(\lambda)}/\Jb$, which corresponds to a global re-scaling of the Hamiltonian.

Note that the Hamiltonian~\eqref{eq:ham_fundamental} has the form
\begin{equation}
    H(\phi, J,\lambda)=\epsilon t J +q\, H_1(J,\phi) \, , 
\end{equation}
for which the equations of motion are
\begin{equation}
    \begin{split}
    \dv{J}{t} & =-q \frac{\partial H_1}{\partial \phi} \\
    \dv{\phi}{t} & =\epsilon t +q\frac{\partial H_1}{\partial J} \, .
    \end{split}
    \label{eq:motion}
\end{equation}
By introducing a new time $\bar t=q \, t$ Eq.~\eqref{eq:motion} can be recast in the following form
\begin{equation}
    \begin{split}
    \dv{J}{\bar t} & = -\frac{\partial H_1}{\partial \phi} \\
    \dv{\phi}{\bar t} & = -\frac{\epsilon}{q^2} \bar t +\frac{\partial H_1}{\partial J} \, .
    \end{split}
\end{equation}
Thus, the small parameter characterizing the adiabaticity is $\bar \epsilon= \epsilon/q^2$, and the new slow time is $\bar \lambda=  \epsilon/q^2 \, \bar t$. We remark that the reasoning can be extended to the case in which $\delta(\lambda)$ is a nonlinear function of $\lambda$, \ie $\delta(\lambda) \approx 1/(2n+1) \left ( \lambda - \lambda_\mathrm{c}\right )^{2n+1}$, where $\lambda_\mathrm{c}$ represents the time of the resonance crossing. This option might be useful in applications in order to improve the overall adiabaticity of the process. In this case, it is easy to show that the small parameter characterizing the adiabaticity is $\bar \epsilon = \epsilon /q^{\frac{2n+2}{2n+1}}$, and the exponent tends to $1$ when $n \to \infty$. 

The Hamiltonian~\eqref{eq:ham_fundamental} is symmetric with respect to the transformation $\tilde J=1-J$ and $\tilde \phi=-\phi$ as indeed
\begin{equation}
\begin{split}
H(\tilde \phi,\tilde J,\lambda) & =\delta(\lambda) (1-\tilde J)-q \sqrt{(1-\tilde J) \tilde J}\sin \tilde \phi \\
& =\delta(\lambda)-H(\phi, J,\lambda)
\end{split}
\end{equation}
so that we have the same dynamics by reverting the time arrow and the behavior for $J\to 0$ is the same as $J\to 1$. 

The level curve that reaches $J=1$ at $\phi=0$ and $\phi=\pi$ is a critical one. It fulfills the equation
\begin{equation}
H(\phi, J,\lambda)=H(0,1,\lambda)
\end{equation}
id est
\begin{equation}
q\sqrt{(1-J)J}\sin\phi - \delta(\lambda)(1-J)=0
\end{equation}
and thus
\begin{equation}
\begin{split}
J(\phi) & =\frac{\delta^2}{\delta^2+q^2 \sin^2 \phi} \to 1 \quad \text{for} \quad \phi \to 0
\end{split}
\end{equation}
that shows how the level curve $J(\phi)$ is tangent to the $J=1$ curve. It is worth stressing that, in spite of being a critical curve, this special level curve of the Hamiltonian $H$ is not a singularity of the dynamics and, in particular, the time spent on this curve is finite (see Appendix~\ref{sec:app1}).

In Fig.~\ref{fig:ham1}, the phase-space portraits of the Hamiltonian~\eqref{eq:ham_fundamental} (assumed to be frozen, \ie with $\lambda$ constant) are shown in the first column, for $q=1$~\footnote{The artificially large value of $q$, together with large values of $\delta$ and of the action, is used to make more visible the key features of the phase-space portrait.} and three values of $\delta$, namely $1, 0, -1$ for the top, center, and bottom plot, respectively.
\begin{figure*}[htb]
\centering
\includegraphics[trim=0truemm 0truemm 10truemm 10truemm, width=.4\textwidth,clip=]{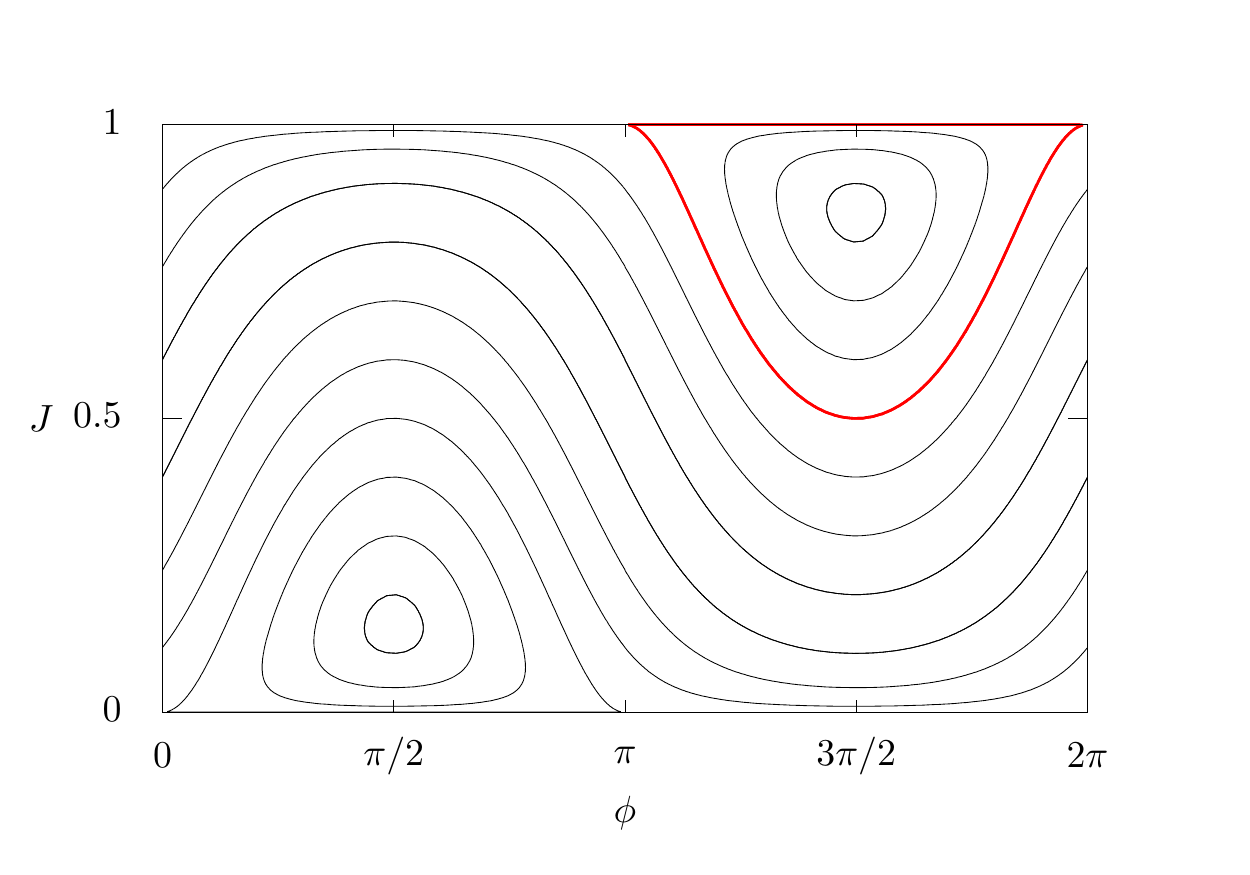}
\includegraphics[trim=0truemm 0truemm 10truemm 10truemm, width=.27\textwidth,clip=]{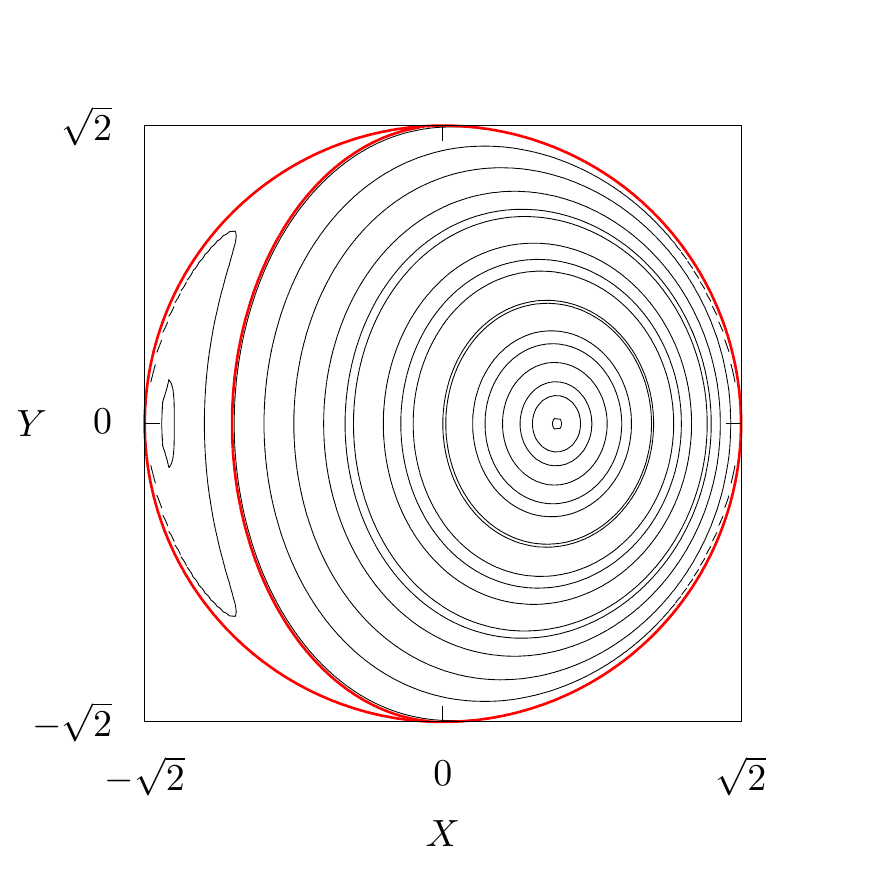}
\includegraphics[trim=20truemm 15truemm 20truemm 20truemm, width=.25\textwidth,clip=]{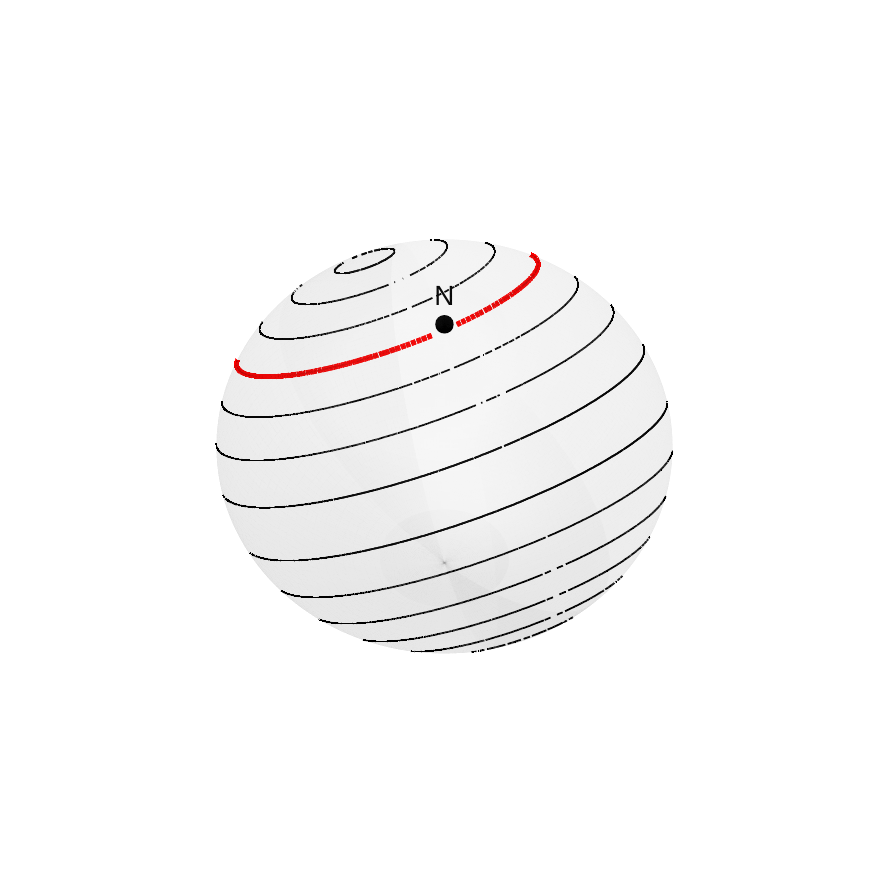}\\
\includegraphics[trim=0truemm 0truemm 10truemm 10truemm, width=.4\textwidth,clip=]{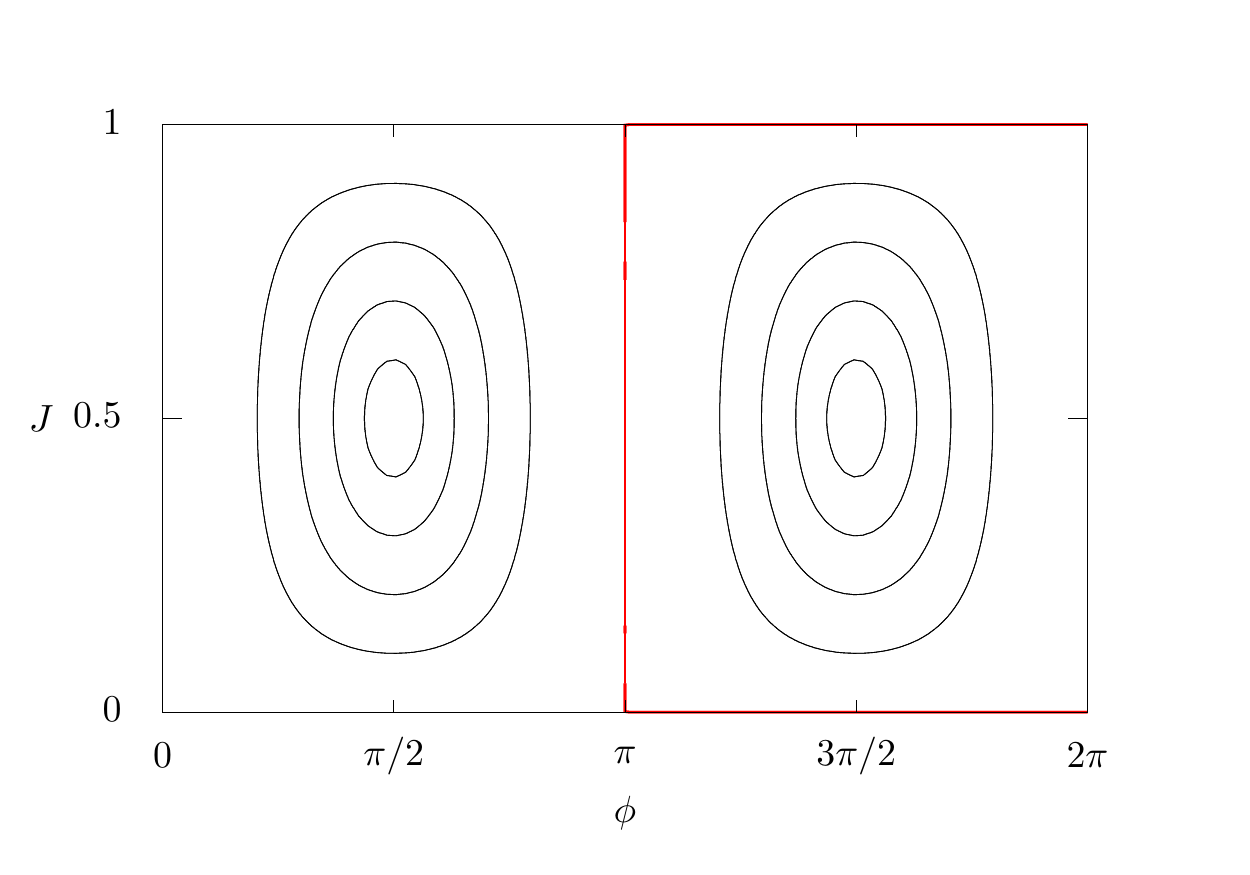}
\includegraphics[trim=0truemm 0truemm 10truemm 10truemm, width=.27\textwidth,clip=]{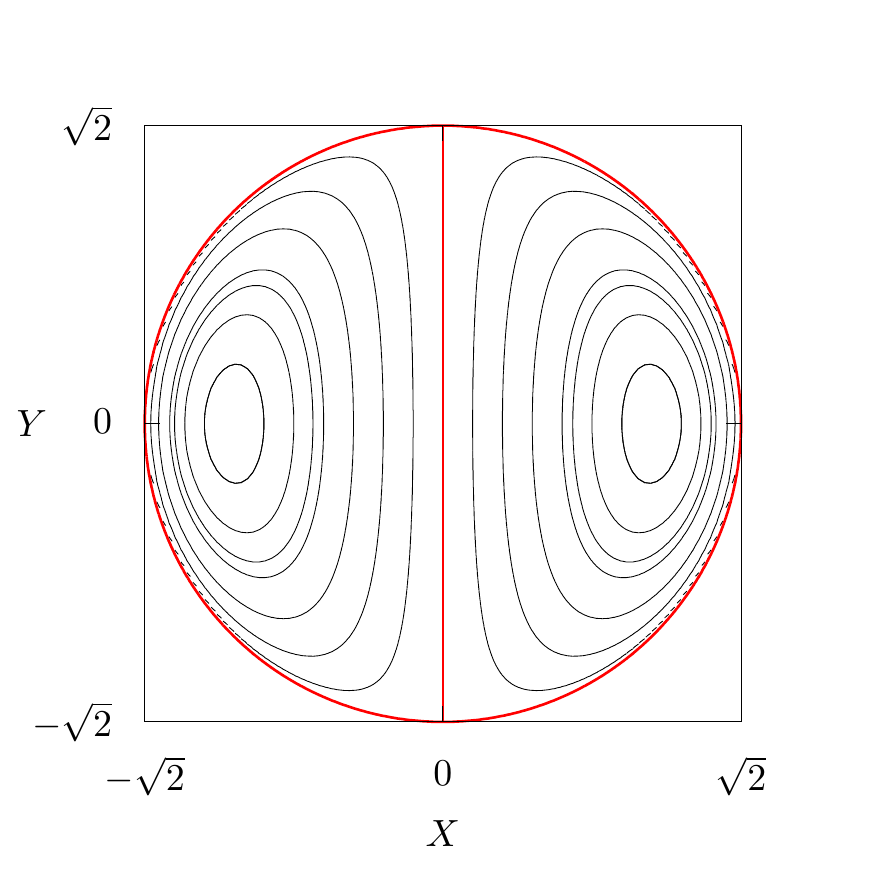}
\includegraphics[trim=20truemm 15truemm 20truemm 20truemm, width=.25\textwidth,clip=]{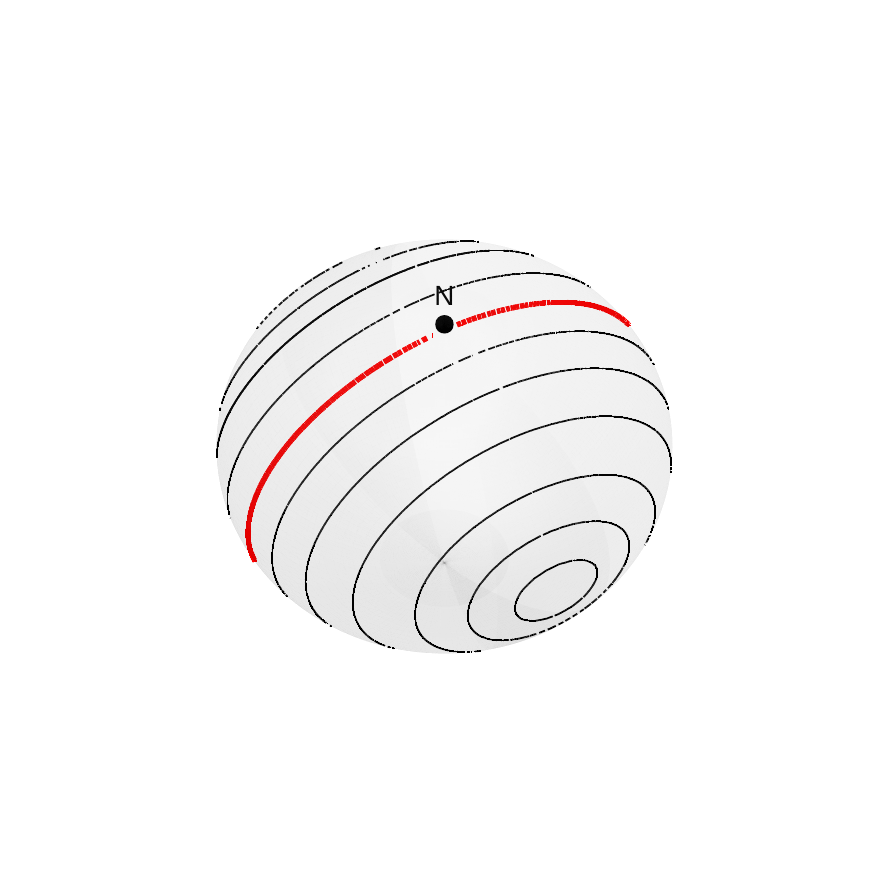}\\
\includegraphics[trim=0truemm 0truemm 10truemm 10truemm, width=.4\textwidth,clip=]{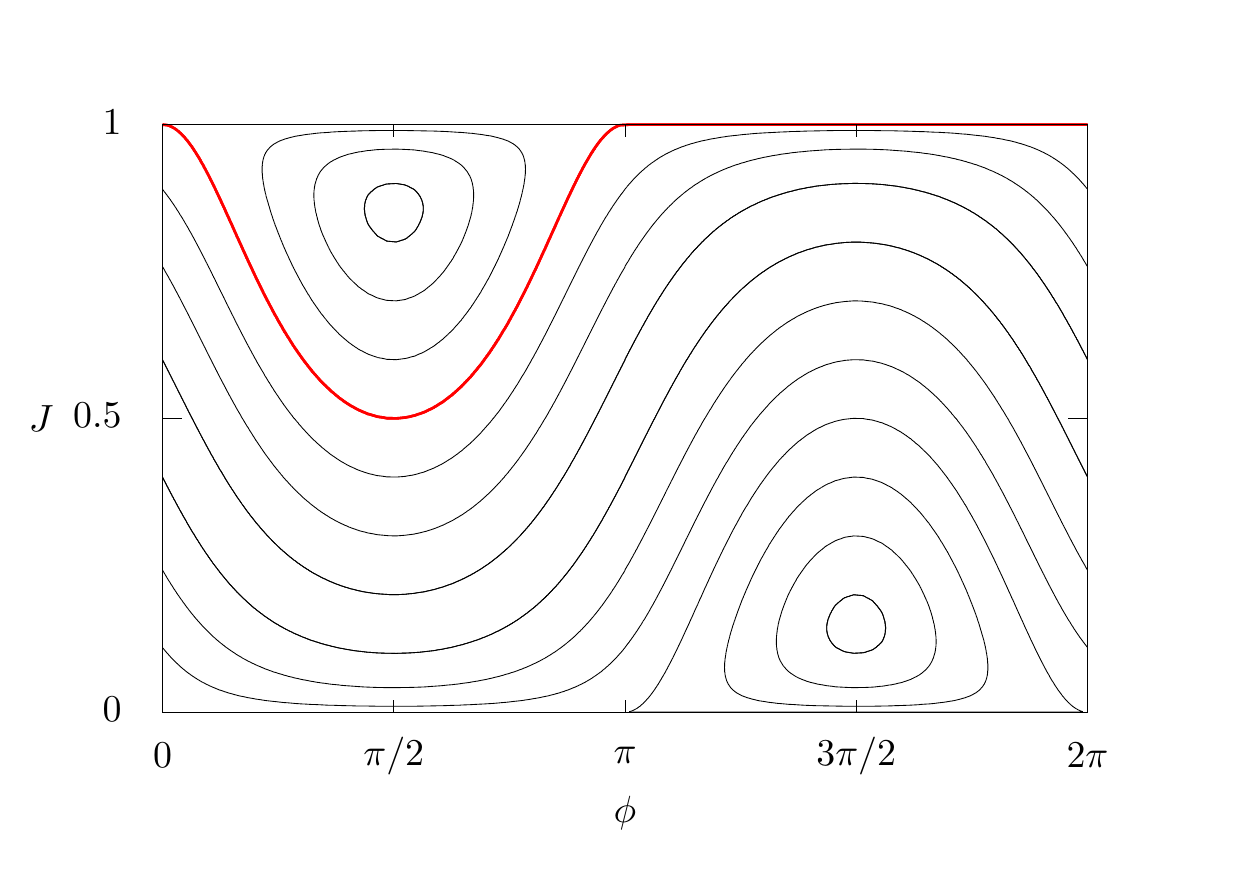}
\includegraphics[trim=0truemm 0truemm 10truemm 10truemm, width=.27\textwidth,clip=]{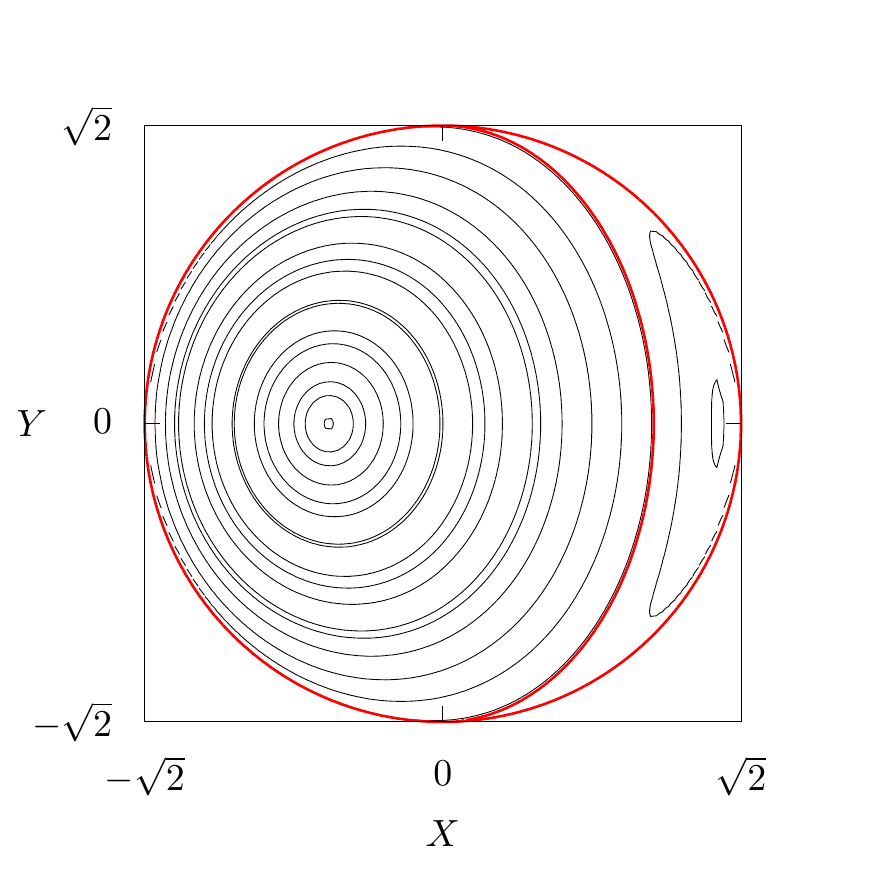}
\includegraphics[trim=20truemm 15truemm 20truemm 20truemm, width=.25\textwidth,clip=]{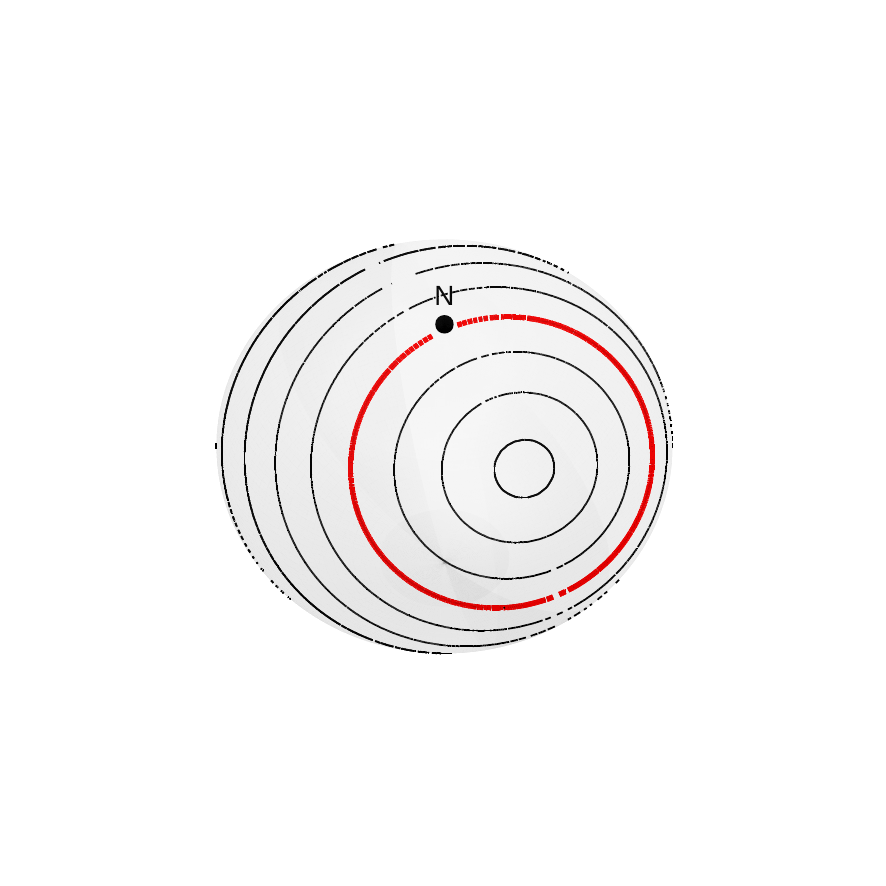}\\
\caption{Phase-space portraits of the Hamiltonian~\eqref{eq:ham_fundamental} for $q=1$ and $\delta=1$ (top), $\delta=0$ (center), and $\delta=-1$ (bottom) in three different representations: action-angle co-ordinates $(\phi,\, J)$ on the left column, Cartesian co-ordinates $(X=\sqrt{2J}\sin\phi,\,Y=\sqrt{2J}\cos\phi)$ in the central column, and on a spherical surface (as described in Section~\ref{sec:sphere}) on the right column, where the coordinate $\phi$ is the polar angle while the azimuthal angle is given by $\theta = \asin(2J-1)$. The red line represents the critical curve (the so-called {\sl coupling arc} in Refs.~\cite{Lee:2651939,PhysRevLett.110.094801}). The plots on the left column are the equirectangular projections of the spherical surface, the poles being represented by the $J=0$ and $J=1$ lines while the plots on the central column are the azimuthal representations centered around the south pole. Note that the artificially-large value of $q$, together with large values of $\delta$ and of $J$, is used to make more visible the key features of the phase-space portrait.}
\label{fig:ham1}
\end{figure*}

The red curves represent the critical curve, which is also called \textsl{coupling arc} in Refs.~\cite{Lee:2651939,PhysRevLett.110.094801}. In the top plot ($\delta=1$), two separated islands are visible, whose areas increase as $\delta$ decreases to zero. Furthermore, there exists a region of \textsl{separatrix curves} around the islands, tangent to the singular lines $J=0$ and $J=1$. When $\delta=0$ (center plot), the islands have maximal area, with a sort of separatrix that connects the singular line through the vertical line $\phi=\pi$. Finally, a symmetric situation when $\delta <0$ is visible in the bottom plot.

The dynamics can also be studied by using the variables
\begin{equation} 
X=\sqrt{2J}\sin\phi \qquad Y=\sqrt{2J}\cos\phi
\label{eq:coordinates}
\end{equation}
so that the Hamiltonian reads as
\begin{equation}
\begin{split}
H(X,Y,\lambda) & =\frac{\delta(\lambda)}{2}(X^2+Y^2)+\\
& + \frac{q}{2}\sqrt{2-(X^2+Y^2)}X \, .
\end{split}
\label{eq:ham_north}
\end{equation}

A limiting circle $X^2+Y^2 = 2$ appears and the dynamics is confined within it, due to the presence of the square root. Moreover, the {\sl coupling arc} is the solution of
\begin{equation}
\delta^2(\lambda)(X^2+Y^2)+q^2X^2 = 2\delta^2(\lambda)
\end{equation}
that separates the accessible domain of phase space into two regions of different size, depending on the value of $\delta(\lambda)$. A sketch of the phase-space portrait is depicted in the central column of Fig.~\ref{fig:ham1}.
\subsection{Visualization of the dynamics on a sphere}\label{sec:sphere}
In the previous section, the dynamics of the Hamiltonian~\eqref{eq:ham_fundamental} has been analyzed by means of two different co-ordinate systems. However, the essential features of the dynamics can be best appreciated by looking at the dynamics generated on a sphere, since we have two singular lines, namely $J=0$ and $J=1$. 

The north pole can be identified with $J=1$ and the south pole with $J=0$. Two charts have to be defined: one describing the southern and one the northern hemisphere, with a non-zero overlap at the equator to provide the necessary compatibility between the two charts. The co-ordinates~\eqref{eq:coordinates} can be used to describe the chart of the northern hemisphere.

There are two symmetrically-located elliptic fixed points on the sphere, whereas the poles are singular lines, \ie the phase dynamics $\phi(t)$ is not defined on the poles, but we observe that the phase velocity $\dot \phi$ increases as one approaches the poles as $1/\sqrt{(1-J)J}$ so that the time spent in the part of the energy-level curves near the lines $J=0$ and $J=1$ tends to $0$. We remark that the level curves tend to be parallel to these singular lines, whereas the angle $\phi$ varies in a neighborhood of $\pi/2$ or $3\pi/2$.

The level line $H(X,Y,\lambda)=0$ close to the origin is given by 
\begin{equation}
\delta(\lambda) (X^2+Y^2)+\sqrt{2} q X=O(3) \, ,
\end{equation}
where $O(3)$ stands for third-order terms in $X$ and $Y$. Such a curve is smooth and represents a circle passing through the origin. In the limit $\delta \to 0$, the radius of curvature diverges and the same is true near the south pole $J=1$, where the variables are given by
\begin{equation}
\tilde X=\sqrt{2(1-J)}\sin\phi \qquad \tilde Y=\sqrt{2(1-J)}\cos\phi
\end{equation}
and the Hamiltonian is
\begin{equation}
\begin{split}
H(\tilde X,\tilde Y,\lambda) & =-\frac{\delta(\lambda)}{2}({\tilde X}^2+{\tilde Y}^2)+\\
& +\frac{q}{2}\sqrt{2-{\tilde X}^2+{\tilde Y}^2}\, {\tilde X}+\delta(\lambda) \, .
\end{split}
\label{eq:ham_south}
\end{equation}
Both Hamiltonians~\eqref{eq:ham_north} and~\eqref{eq:ham_south} are analytic at their origin, and they have a singularity at
\begin{equation}
\frac{{\tilde X}^2+{\tilde Y}^2}{2}=\frac{{X}^2+{Y}^2}{2}=1
\end{equation}
corresponding to the points $J=1$ and $J=0$, respectively. However, the singularity is not in the dynamics, but only in the co-ordinates. Hence, the dynamics on the sphere has no singularity, although it cannot be described by a single  chart and this is an essential point for our analyses. The dynamics on the sphere is represented in the plots on third column of Fig.~\ref{fig:ham1}.

When the time dependence is considered, and in particular the limit $\delta(\lambda) \to 0$ is analyzed, then one can introduce the action-angle variables $(\theta, \, I)$ for each chart of the 1DoF frozen Hamiltonian~\eqref{eq:ham_north}.
\subsection{Analysis of the pseudo-resonance-crossing process}
Let us assume that  $\delta(\lambda)=\delta_{\rm max} \, \epsilon t$
where $t\in [-\epsilon^{-1},\epsilon^{-1}]$ and $\delta(\lambda) \in [-\delta_{\rm max}, \delta_{\rm max}]$. 

The fixed points of the Hamiltonian~\eqref{eq:ham_fundamental} correspond to $\phi_x^\ast=\pi/2,3\pi/2$
and their action $J^\ast$ is given by 
\begin{equation}
    \frac{2\delta_{\rm max}}{q} \lambda=\pm\frac{1-2J^\ast}{\sqrt{J^\ast(1-J^\ast)}} \, ,
\end{equation}
where the plus sign refers to $\phi^\ast=\pi/2$.

The applicability of adiabatic theory at a resonance crossing relies on the control of a slow phase change during the variation of the parameter $\delta(\lambda)$. When $\lambda=0$, $J^\ast=1/2$ and 
\begin{equation}
    \dv{J^\ast(0)}{t} \simeq -\frac{\delta_{\rm max}}{q}\epsilon
\end{equation}
and this quantity has to be small, \ie $\ll 1$, to apply adiabatic theory. 

The two fixed points move to opposite directions during the resonance-crossing stage, and the corresponding resonance islands have an amplitude estimated by
\begin{equation}
    \Delta J(\lambda)=\frac{1}{\left (\frac{\delta_{\rm max}}{q}\right )^2 + 1} \, .
\end{equation}

$\Delta J(\lambda)$ is maximum when $\delta(\lambda)=0$, but at the boundary values, \ie $\pm \delta_{\rm max}$, it can be small if $\delta_{\rm max}/q \gg 1$. In such a case, trapping inside the island can occur near one of the borders, whereas de-trapping occurs near the other one and such a phenomenon happens in a symmetric fashion with respect to the horizontal line $1/2$, so that the values $I_x$ and $I_y$ (\ie the Courant-Snyder invariants and then also the beam emittances) are exchanged by keeping approximately fixed their sum. 
\section{Analysis of the dynamics using the normal modes} \label{sec:digression}
By following the approach described in detail in Appendix~\ref{sec:app2}, the prototype Hamiltonian to study the emittance exchange process can be written in the normal modes space in the following form
\begin{equation}
H(\phi, J,\lambda)=\gamma(\lambda)J+\epsilon\sqrt{(1-J)J}\sin\phi \, ,
\label{eq:ham_fundamental1}
\end{equation}
where, without loss of generality, we assume $J_2=1$, so that $J=0$ and $J=1$ are singular lines for the Hamiltonian. Although  the geometrical properties of the dynamics generated by the Hamiltonian~\eqref{eq:ham_fundamental1} coincide with those of~\eqref{eq:ham_fundamental}, the two descriptions are carried out in different spaces, namely that of physical co-ordinates for Eq.~\eqref{eq:ham_fundamental} or that of the normal modes for Eq.~\eqref{eq:ham_fundamental1}. Hence, the exchange of the invariants occurs in different spaces. 

The Hamiltonian~\eqref{eq:ham_fundamental1} allows to study the effect of the adiabaticity in the frequency modulation on the preservation of the action variables $(J_1, J_2)$ and of their approximations $(\Ja,\Jb)$ in the physical planes. 

The Hamiltonian~\eqref{eq:ham_fundamental1}, being of the same form as~\eqref{eq:ham_fundamental}, has  two elliptic fixed points at $\phi^\ast=\pi/2,\,3\pi/2$ and
\begin{equation}
\gamma(\lambda)=\pm \epsilon\frac{1-2J^\ast}{\sqrt{J^\ast(1-J^\ast)}}
\end{equation}
so that if $\gamma\gg \epsilon$ they are very close to boundaries $J=0$ and $J=1$, which means that the resonance trapping is not possible, and we have simply the invariance of $J$ for $\epsilon \to 0$. The maximum resonance amplitude occurs at the minimum value of $\gamma$
\begin{equation}
\Delta J=\frac{1}{O \left ( \left (\displaystyle{\frac{q}{\epsilon}} \right )^2 \right )+1}
\end{equation}
so that $\Delta J=O(1)$ only if $q=O(\epsilon)$ whereas if $q\gg \epsilon$, then $\Delta J$ is negligible, which clearly describes the interplay between the two small parameters $q$ and $\epsilon$. 

Even for $q=O(\epsilon)$, $I_1$ and $I_2$ are adiabatic invariants, which are referred to the phase planes defined by the eigenvectors. At the beginning of the crossing process, given that $q$ is small, the two planes are close to the original phase planes $(X,P_X)$ and $(Y,P_Y)$, and the same holds for the initial actions so that $I_1 \simeq I_x(0)$ and $I_2\simeq I_y(0)$. With an error $O(q)$, at the end of the process the two planes are exchanged, and the same is true for the emittances with the same approximation $O(q)$. Hence, $q$ defines the maximum possible emittance exchange whereas $\epsilon\ll 1$ allows a conservation of the adiabatic invariants.

Furthermore, the action-angle variables are analytic for $\gamma(\lambda) \to 0$ and the Hamiltonian reduces to the form
\begin{equation}
H(\phi, J,\lambda)=H(J,\lambda)+\epsilon H_1(\phi, J,\lambda)
\end{equation}
and is analytic on the sphere. It is then possible to apply the Theorem reported in Ref.~\cite{NEISHTADT198158} to the Hamiltonian $H(\phi, J,\lambda)$ to state that the change of the action for a given orbit of the system is exponentially-small, \ie $\Delta J = O (\exp(-c/\epsilon))$ with $c$ a positive constant, when  $\lambda$ varies, which corresponds to the crossing of the original difference resonance.

It is worth stressing that the same remarks made for the Hamiltonian~\eqref{eq:ham_fundamental} about the re-scaled adiabaticity parameter hold also for the Hamiltonian~\eqref{eq:ham_fundamental1}. Therefore, one can state that $\Delta J = O (\exp(-c \, q^2/\epsilon))$ in case of a resonance crossing linear in $\lambda$, or $\Delta J = O (\exp(-c q^\frac{2n+2}{2n+1}/\epsilon))$ in case of a nonlinear crossing of the resonance. Note that a nonlinear resonance crossing is more advantageous in terms of adiabaticity of the process with respect to a linear one.
\section{Impact of detuning with amplitude}\label{sec:detuning}
In the presence of detuning with amplitude generated by nonlinearities, the dynamics is governed by the Hamiltonian~\eqref{eq:hamq} plus the terms~\cite{PhysRevE.49.2347} 
\begin{equation}
\begin{split}
    H_\text{det}(p_x, p_y, x, y) &= \alpha_{xx}\qty(\frac{x^2 + p_x^2}{2})^2 + \\ 
    &+ 2\alpha_{xy}\qty(\frac{x^2 + p_x^2}{2})\qty(\frac{ y^2 + p_y^2}{2})+ \\
    &+\alpha_{yy}\qty(\frac{ y^2 + p_y^2}{2})^2 \, .
\end{split}
\end{equation}

This approach assumes that nonlinear resonances are not excited by the nonlinearities or that those resonances are far from the difference resonance we are considering. By performing the same substitutions that led to the Hamiltonian~\eqref{eq:ham_fundamental} starting from~\eqref{eq:hamq}, $H_\text{det}$ becomes
\begin{equation}
H_\text{det}(\phi_\text{a}, \Ja) = \alpha_{\rm aa}\Ja^2 + \alpha_{\rm ab}\Ja\Jb \, ,
\end{equation}
where a constant term in $J_\mathrm{b}$ has been discarded as $J_{\rm b}$ is a constant of motion and hence the constant term is irrelevant for the dynamics of $J_{\rm a}$ and $\phi_{\rm a}$, and
\begin{equation}
\alpha_\mathrm{aa} = \alpha_{xx}-2\alpha_{xy}-\alpha_{yy} \qquad \alpha_\mathrm{ab} = 2\alpha_{xy}-\alpha_{yy}
\end{equation}
and the complete Hamiltonian becomes
\begin{equation}
\begin{split}
H(\phi_\text{a}, J_\mathrm{a},\lambda) & = (\delta(\lambda) +  \alpha_\text{ab}\Jb)J_\mathrm{a} + \alpha_\text{aa} \sqrt{\omega_x(\lambda) \omega_y(\lambda)} \, J_\mathrm{a}^2 +\\ & + q\sqrt{J_\mathrm{a}(J_\mathrm{b}-J_\mathrm{a})}\cos\phi \, .
\end{split}   
\label{eq:ham_det}
\end{equation}

It is obvious that $\alpha_{\rm ab}$ can be reabsorbed in the definition of $\delta$, which effectively would correspond to shifting the resonant condition to $\omega_x-\omega_y + \alpha_{\rm ab}\Jb = 0$ or, equivalently, to shifting the time at which the resonance is crossed. It is also evident that acting on the three physical quantities $\alpha_{xx},\,\alpha_{xy},\,\alpha_{yy}$ it is possible to control the values of $\alpha_\mathrm{aa}$ and $\alpha_\mathrm{ab}$ independently on each other. As it was done for the Hamiltonian~\eqref{eq:ham_fundamental}, it is possible to shift the phase of $\phi$ and set $J_{\rm b}=1$ to cast~\eqref{eq:ham_det} in the following form
\begin{equation}
H(\phi, J,\lambda) = \delta(\lambda) J + \alpha(\lambda) J^2 + q\sqrt{J(1-J)}\sin \phi \, .
\label{eq:ham_det1}
\end{equation}

Whenever the analysis would be carried out in the normal modes' co-ordinates, then it would be immediate to find that the detuning with amplitude would lead to the following general Hamiltonian
\begin{equation}
H(\phi, J,\lambda) = \gamma(\lambda) J + \hat \alpha(\lambda) J^2 + \epsilon \sqrt{J(1-J)}\sin \phi \, ,
\label{eq:ham_det2}
\end{equation}
where also in this case $\gamma(\lambda)$ incorporates a constant term $\alpha_{\rm ab}$ with respect to the original definition used in~\eqref{eq:ham_fundamental1}.

\begin{figure*}
    \centering
    \includegraphics[trim=0truemm 0truemm 10truemm 10truemm,width=.32\textwidth,clip=]{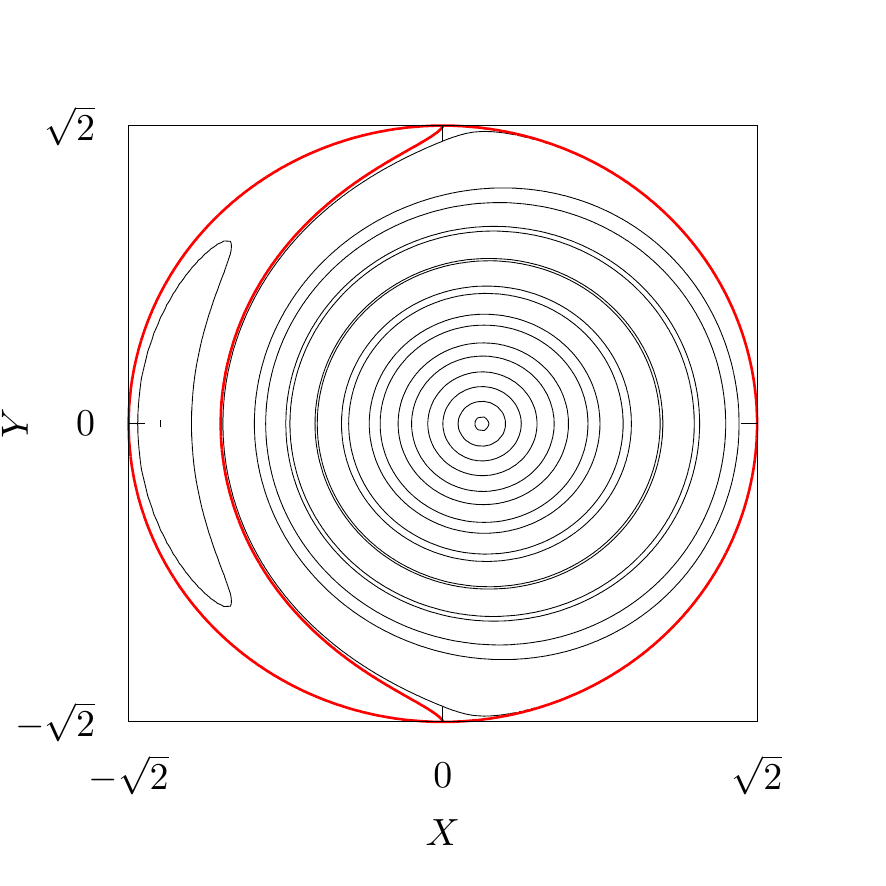}
    \includegraphics[trim=0truemm 0truemm 10truemm 10truemm,width=.32\textwidth,clip=]{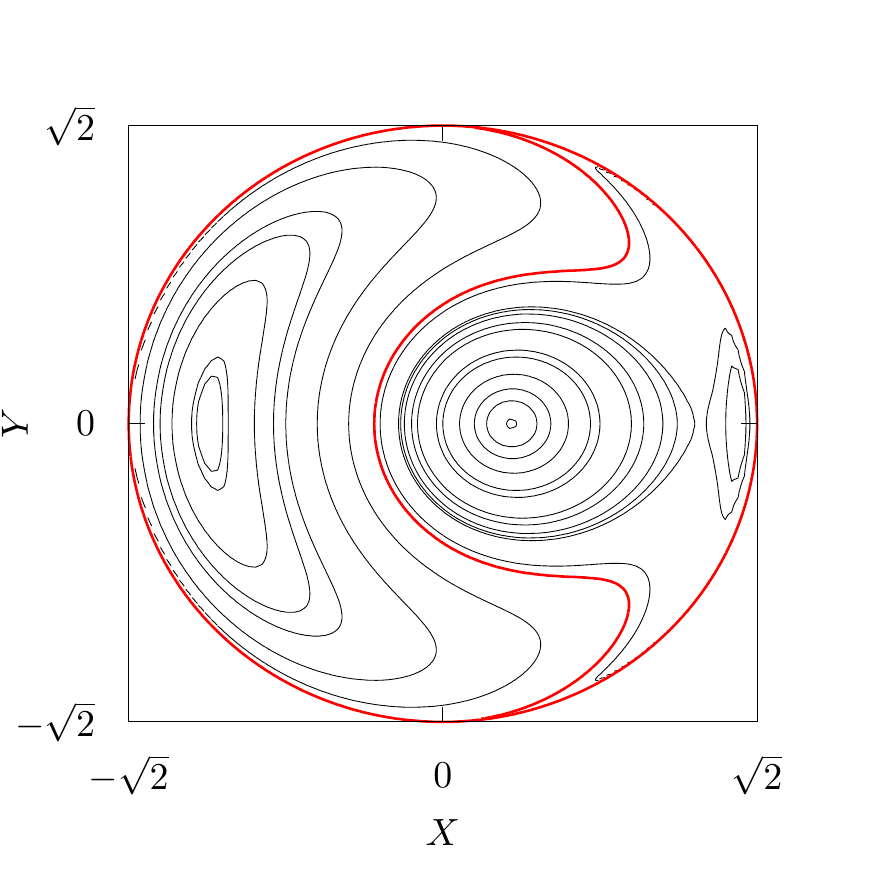}
    \includegraphics[trim=0truemm 0truemm 10truemm 10truemm,width=.32\textwidth,clip=]{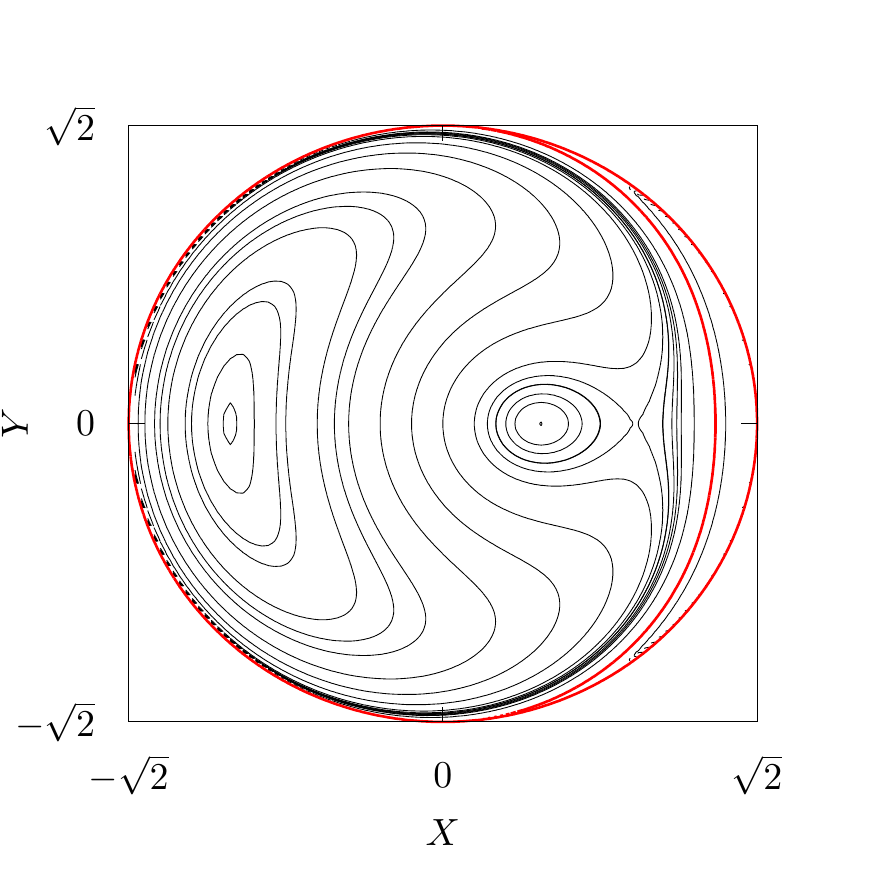}
    \caption{Phase portraits of the  Hamiltonian~\eqref{eq:mapfinal_oct} with $\alpha_{\rm aa}=1$, $\alpha_{\rm ab}=0$, $q=0.25$, $\Jb=1$ at $\delta=-1$ (left), $\delta=-0.58$ (center), $\delta=-0.44$ (right), in Cartesian $(X,Y)$ co-ordinates. The orbits that pass through $X=0$, $Y=\pm\sqrt{2}$ are shown in red.}
    \label{fig:phsp_oct}
\end{figure*}

The parameter $\alpha(\lambda)$ (or $\hat \alpha(\lambda)$) has a fundamental impact on the phase-space topology as, when $\alpha=0$ the Hamiltonians~\eqref{eq:ham_fundamental} or~\eqref{eq:ham_fundamental1} have only two elliptic fixed points, whereas when $\alpha \neq 0$ an additional pair of one elliptic and one hyperbolic fixed point might be generated (see Ref.~\cite{PhysRevE.49.2347}). The conditions for the existence of these additional fixed points are discussed in Appendix~\ref{sec:app3}. Indeed, the presence of a hyperbolic fixed point implies the existence of a separatrix, which introduces a singularity in the phase-space structure and hence alters the character of the dynamics. Examples of phase-space portraits for different values of $\delta$ are shown in Fig.~\ref{fig:phsp_oct}, and the hyperbolic fixed points are clearly visible.

In particular, the nice property about the exponentially-small change of $J$, linked to the analyticity of the dynamics of~\eqref{eq:ham_fundamental1}, is lost. From the discussion presented in the Appendix~\ref{sec:app3} it follows that when $\alpha_{\rm aa}$ is sufficiently small, no extra fixed point is present and the phase-space topology is unchanged with respect to~\eqref{eq:ham_fundamental} with no separatrix present and hence exponentially-small bound on the variation of the invariant change during the resonance-crossing process.
\section{Digression: two-way crossing of the coupling resonance} \label{sec:digr}
So far, the focus has been on the analysis of the adiabaticity properties of the crossing of the coupling resonance for a linear and nonlinear system. However, another process is possible and is interesting to consider, namely the two-way crossing of the resonance. Such a process would allow studying the reversibility of the resonance-crossing process and, in particular, the impact on the phase, as that on the action variable is already fully covered by the considerations made in the previous sections. The treatment proposed in~\cite{Arnold:937549} is used to deal with the two-way resonance crossing. The starting point is the Hamiltonian~\eqref{eq:ham_det2} in which the parameter $\gamma(\lambda)$ is supposed to describe a closed curve when $\lambda \in [0,1]$, corresponding to $t \in [0, 1/\epsilon]$. The change in the phase of the action-angle variables when the system moves along the closed curve can be evaluated by~\cite{Arnold:937549} 
\begin{equation}
    \phi \left (\frac{1}{\epsilon} \right )-\phi(0)= \chi_\mathrm{dyn}+\chi_\mathrm{geom}+\chi_\mathrm{rem} \, ,
\end{equation}
where the three terms can be generically computed, assuming the Hamiltonian is given as $H=H_0(J,\gamma(\tau))+\epsilon H_1(\phi, J,\tau)$, according to 
\begin{equation}
    \begin{split}
        \chi_\mathrm{dyn} & = \frac{1}{\epsilon} \int_0^1 \frac{\partial H_0(J(\frac{\tau}{\epsilon}),\gamma(\tau))}{\partial J} \, \dd \tau\\
        \chi_\mathrm{geom} & = \int_0^1 \frac{\partial \mathcal{H}_1(J(0),\tau)}{\partial J} \, \dd \tau \, , \quad \mathcal{H}_1=\langle H_1\rangle_\phi\\
        \chi_\mathrm{rem} & = \epsilon \int_0^{1/\epsilon} \frac{\partial H_1(\phi(\eta), J(\eta),\epsilon \, \eta)}{\partial J} \, \dd \eta - \chi_\mathrm{geom} \, .
    \end{split}
\end{equation}
The first term, $\chi_\mathrm{dyn}$, takes into account the dynamical change of the phase and is relevant in the adiabatic regime, \ie when $\epsilon \ll 1$ and depends only on the form of $H_0$. The second term, $\chi_\mathrm{geom}$, depends on the angular average of $H_1$ and is the so-called Berry phase~\cite{Berry:1984,Berry:1985} or Hannay angle~\cite{Hannay:1985}. The third term, $\chi_\mathrm{rem}$, is relevant in the non-adiabatic regime and depends only on $H_1$.

These general definitions can be specialized to the case of the Hamiltonian~\eqref{eq:ham_det2} and one obtains
\begin{equation}
    \begin{split}
        \chi_\mathrm{dyn} & = \frac{1}{\epsilon} \int_0^1 \gamma(\tau) \, \dd \tau\\
        \chi_\mathrm{geom} & = 0 \\
        \chi_\mathrm{rem} & = \frac{\epsilon}{2} \int_0^{1/\epsilon} \frac{1-2J(\eta)}{\sqrt{\left (1-J(\eta) \right ) J(\eta) }} \sin \phi(\eta)\, \dd \eta \, ,
    \end{split}
\end{equation}
where $\chi_\mathrm{geom}=0$ is due to the special form of $H_1$, which is zero when averaged over the angle $\phi$. Note that $\chi_\mathrm{dyn}$ is independent on the action variable, whereas it depends on $\epsilon$. It is worth noting that geometrically, $\chi_\mathrm{dyn}$ represents the area enclosed by the closed curve described by $\gamma (\tau)$, and such an area is zero in case resonance is crossed in the same way in each of the two directions.  $\chi_\mathrm{rem}$ depends on the action-angle variables. This means that in the adiabatic regime the phase is affected by an $\epsilon$-dependent shift, only, whereas in the non-adiabatic case the shift depends also on the action-angle variables. Therefore, in the adiabatic case the distribution of initial conditions is rigidly rotated, \ie by an amplitude-independent angle, in a two-way crossing of the coupling resonance (the action variable being only very weakly affected), while in the non-adiabatic case the initial distribution undergoes a nonlinear deformation by a two-way resonance-crossing process. 

It is worth stressing that the rigid rotation of the initial distribution in case of a two-way crossing of the coupling resonance in the adiabatic regime is a consequence of the special form of $H_0$, which is linear in the action variable. Indeed, whenever detuning with amplitude is considered, $H_0$ is no longer a linear function of $J$ and, hence, even in the adiabatic regime, the initial distribution will be rotated by an amplitude-dependent quantity. This means that a two-way crossing of the resonance is never a fully reversible process for a nonlinear system. Hence, a periodic crossing of the coupling resonance, even if it occurs adiabatically, will always distort the distribution, with an adverse effect on the preservation of the linear invariants. In physical terms, this is the situation for a circular accelerator operated with non-zero chromaticity and close to the coupling resonance, which can be crossed due to the tune modulation induced by the chromaticity.
\section{The map model} \label{sec:map}
A system made of a FODO cell and a skew quadrupole, which is interpolated by the phase flow of the Hamiltonian~\eqref{eq:hamq}, is described by the one-turn map given by
\begin{equation}
    \bm{x}_{n+1} = \bm{M}_\text{FODO}\bm{M}_\text{Skew} \, \bm{x}_n
\label{eq:one-turn}
\end{equation}
where
\begin{equation}
    \bm{M}_\text{Skew}=\begin{pmatrix}1 & 0 & 0 & 0 \\ 0 & 1 & \hat{q} & 0 \\ 0 & 0 & 1 & 0 \\ \hat{q} & 0 & 0 & 1\end{pmatrix}
\end{equation}
Using the steps detailed in the Appendix~\ref{sec:app4}, it is possible to recast Eq.~\eqref{eq:one-turn} as a H\'enon-like map as
\begin{equation}
\begin{pmatrix} X\\X'\\ Y\\ Y' \end{pmatrix}_{n+1} =  \begin{pmatrix} \block(2,2) {\bm{R}(\omega_x)} &  \block(2,2) {\bm{0}} \\ & & & \\ \block(2,2) {\bm{0}} & \block(2,2) {\bm{R}(\omega_y)} \\ & & & \end{pmatrix}\setlength\arraycolsep{0pt} \begin{pmatrix*}[r] X & \\ X' &+q Y\\ Y &\\Y'&+q X \end{pmatrix*}_n
\label{eq:mapfinal}
\end{equation}
and this is the map used in the numerical simulations presented in Section~\ref{sec:res}. 

The analysis of the impact of the detuning with amplitude has been studied by modifying the map~\eqref{eq:mapfinal} including the effect of an octupole in the single-kick approximation~\cite{Bazzani:262179}, whose strength is defined as
\begin{equation}
    K_3=\frac{\ell}{B_0\rho} \frac{\partial^3 B_y}{\partial x^3} \, ,
\end{equation}
where $B_0\rho$ is magnetic rigidity, $\ell$ is the magnetic length, and $B_y$ is the vertical component of the magnetic field. The nonlinear kick generated by a single octupole can be easily determined~\cite{,PhysRevSTAB.7.024001} in the co-ordinate system used to construct the transfer map and the final result reads
\begin{equation}
\begin{split}
 \begin{pmatrix} X\\X'\\ Y\\ Y' \end{pmatrix}_{n+1} & =  \begin{pmatrix} \block(2,2) {\bm{R}(\omega_x)} &  \block(2,2) {\bm{0}} \\ & & & \\ \block(2,2) {\bm{0}} & \block(2,2) {\bm{R}(\omega_y)} \\ & & & \end{pmatrix}\setlength\arraycolsep{0pt} \cdot \\ & \cdot \begin{pmatrix*}[r] X & & \\ X' & +q Y & +\frac{K_3}{6} \beta_x^2 (X^3-3\chi X Y^2)\\ Y & & \\Y'&+q X &-\frac{K_3}{6} \beta_x^2 (\chi^2 Y^3-3\chi X^2 Y)\end{pmatrix*}_n \, ,
\end{split}
\label{eq:mapfinal_oct}
\end{equation}
where $\chi=\beta_y/\beta_x$. Note that in all simulations the following has been chosen $\beta_x=\beta_y=1=\chi$. The link between the map~\eqref{eq:mapfinal_oct} and the parameters $\alpha_{xx}, \alpha_{xy}, \alpha_{yy}$ can be established using Normal Forms~\cite{Bazzani:262179} and it is possible to show that
\begin{align}
    \alpha_{xx} & = - \frac{K_
    3}{32} \beta_x^2 \nonumber \\
    \alpha_{xy} & = \phantom{-} \frac{K_
    3}{8} \beta_x \beta_y \\
    \alpha_{yy} & = - \frac{K_
    3}{32} \beta_y^2 \nonumber
\end{align}
and this enables to build a perfect correspondence between the map and the Hamiltonian model.
\section{Results of numerical simulations} \label{sec:res}
Numerical simulations have been performed using both the Hamiltonian and the map models. However, based on the fact that the results obtained are the same, only those obtained with the map will be presented here. 
\subsection{Linear map model}
As a first step, numerical simulations have been carried out to evaluate the dependence of the emittance-exchange phenomenon on the adiabaticity of the resonance-crossing process. In the simulations, $\omega_y$ has been changed while keeping $\omega_x$ constant. Thus, $\delta(\lambda) = \omega_x - \omega_y(\lambda)$ is changed from a negative to a positive value passing through zero. As a figure of merit, we used the function $P_\mathrm{na}$, introduced in~\cite{PhysRevAccelBeams.23.044003}, which is defined as
\begin{equation}
    P_\mathrm{na} = 1 - \frac{\av{I_{x,\text{f}}} - \av{I_{x,\text{i}}}}{\av{I_{y,\text{i}}} - \av{I_{x,\text{i}}}} \, ,
\end{equation}
where $I_{z,\text{i}}$ and $I_{z,\text{f}}$ are the initial and final linear action variables, respectively. Therefore, $P_\text{na}$, is zero when a perfect exchange is attained and one when no exchange occurs.

\begin{figure}[htb]
    \centering
    \includegraphics[width=0.7\textwidth,clip=]{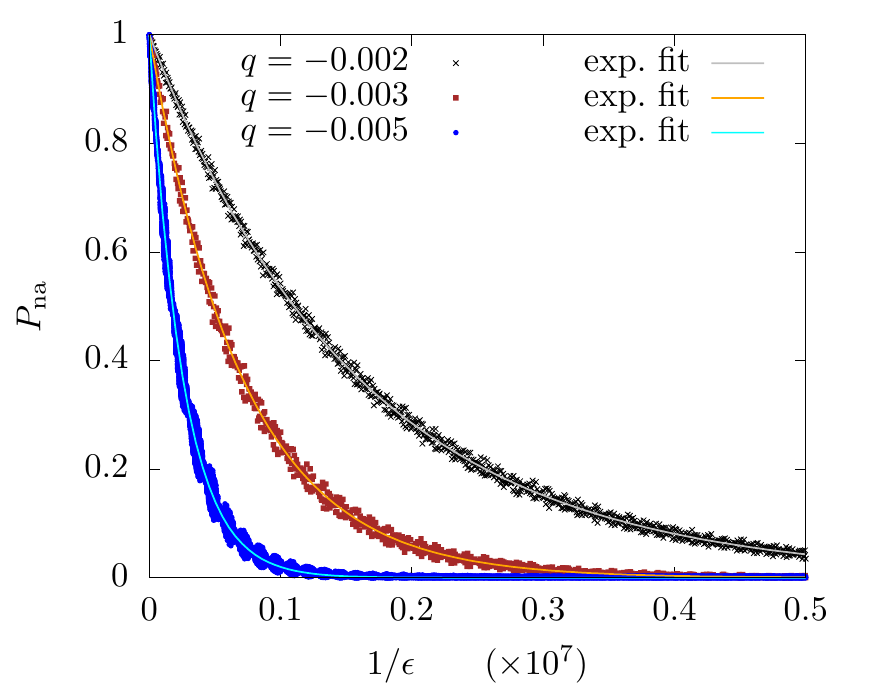}
    \caption{Evolution of $P_\text{na}$ as a function of $1/\epsilon$ for an exponential distribution of initial conditions, and different values of $q$. Exponential fits are also presented. The map~\eqref{eq:mapfinal} has been used, with parameters $\omega_x = 2.602,\, \omega_{y, \rm i}=2.5, \,\omega_{y, \rm f}= 2.7$, and a set of initial conditions with $\av{I_{x, \rm i}}=10^{-4}$, $\av{I_{y, \rm i}}=4\times 10^{-4}$.}
    \label{fig:varnum}
\end{figure}

The evolution of a set of initial conditions, representing a beam exponentially distributed in $I_x$, \ie $\rho(I_x) = ({N_0}/\av{I_x})\exp(-I_x/\av{I_x})$, has been computed by means of the map~\eqref{eq:mapfinal}, while varying $\omega_y$ in the fixed interval $\omega_{y, \rm i}=2.5$ and $\omega_{y, \rm f}=2.7$ over a given time interval $N$. According to the Hamiltonian theory presented in the previous sections, we expect that $\av{I_x}$ becomes $\av{I_y}$ after the resonance crossing. What we observe in Fig.~\ref{fig:varnum} is a clear exponential dependence of $P_\text{na}$ as a function of $1/\epsilon$, in evident agreement with the findings of Ref.~\cite{PhysRevAccelBeams.23.044003},  and also in perfect agreement with the discussion carried out previously. 

The exponential behavior of $P_\text{na}$ features a clear dependence on $q$. However, an oscillatory behavior is also observed due to the neglected terms $O(q^2)$. We have been studying this effect by means of dedicated numerical simulations, in which the evolution of $P_\text{na}$ at large number of turns has been probed. 

\begin{figure*}[htb]
\centering
    \includegraphics[trim=-8truemm 0truemm 6truemm 2truemm, width=.49\textwidth,clip=]{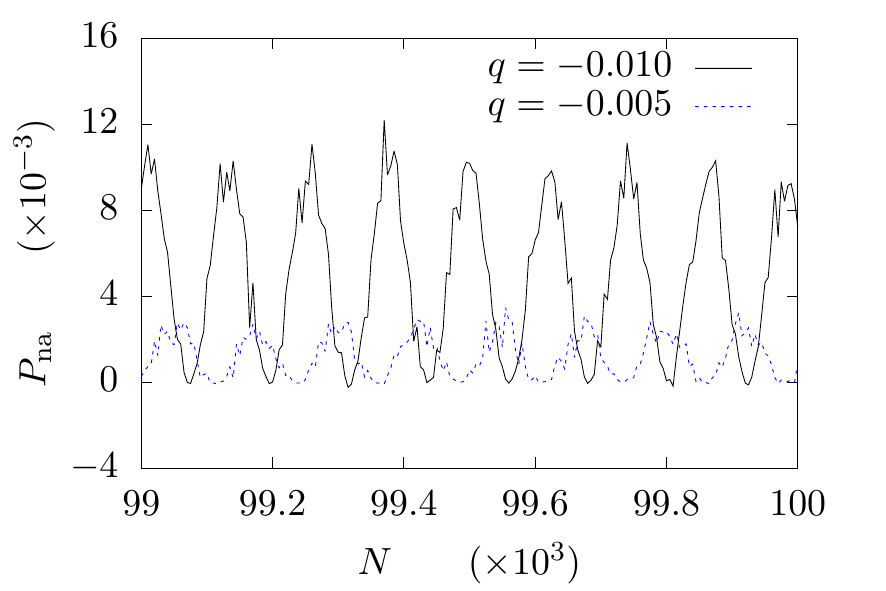}
    \includegraphics[trim=-2truemm 0truemm 6truemm 2truemm,width=.49\textwidth,clip=]{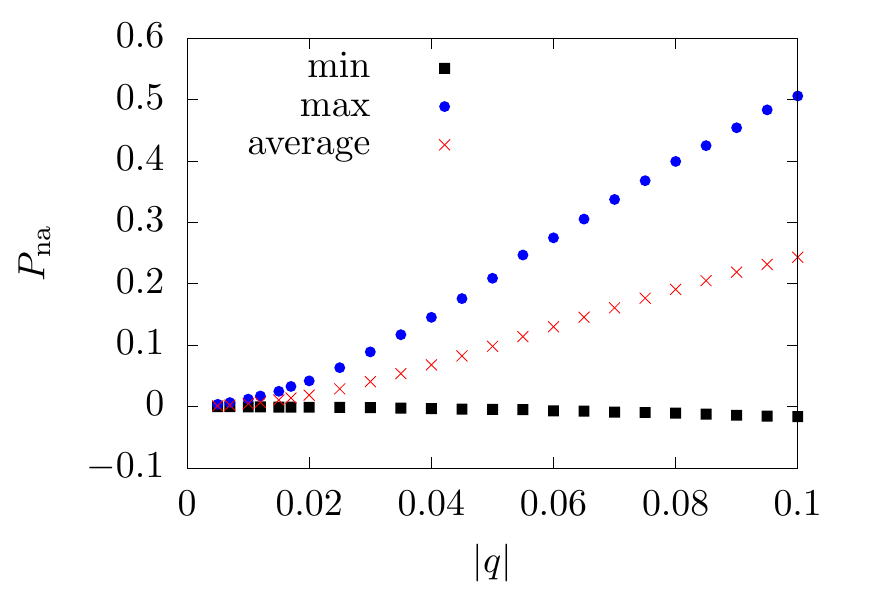}
    \includegraphics[trim=1truemm 0truemm 6truemm 2truemm,width=.49\textwidth,clip=]{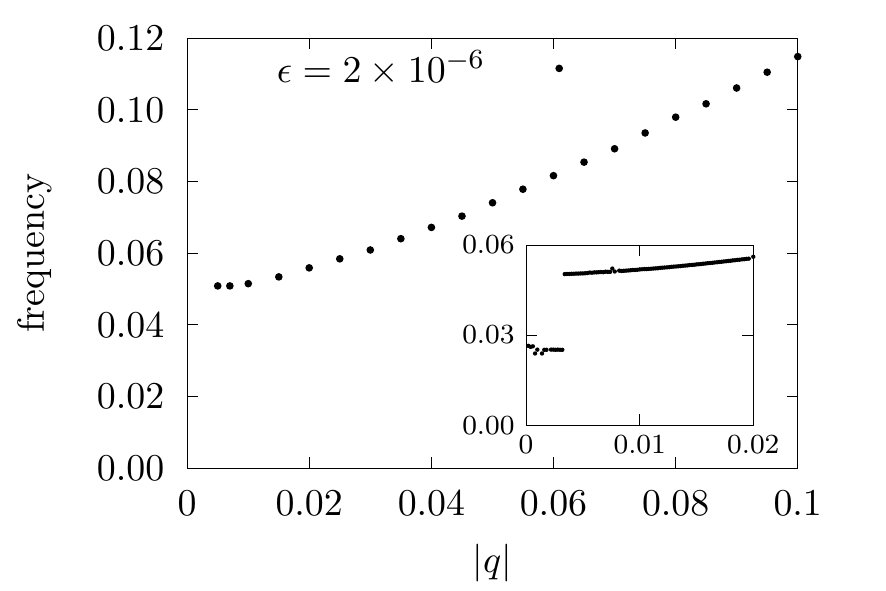}
    \includegraphics[trim=-2truemm 0truemm 6truemm 2truemm,width=.49\textwidth,clip=]{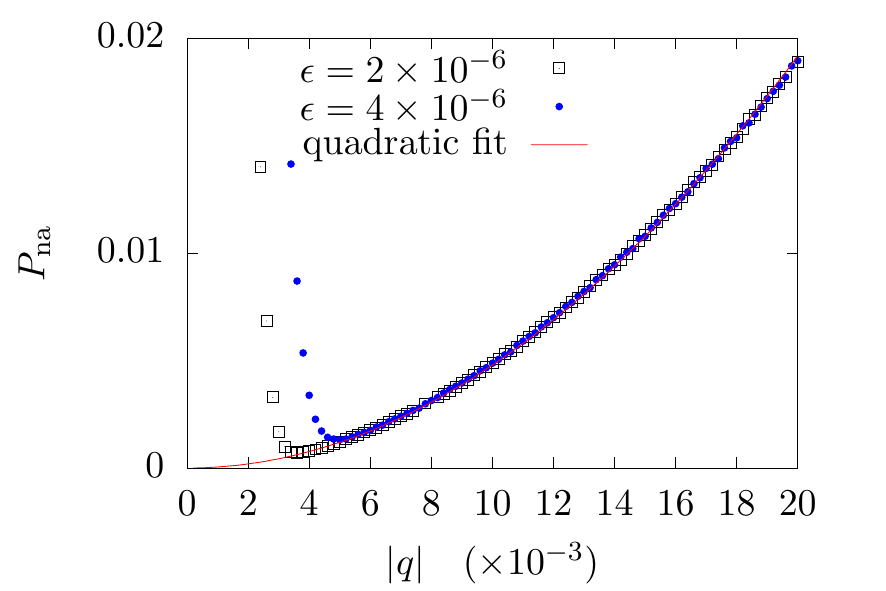}
    \includegraphics[trim=-2truemm 0truemm 6truemm 2truemm,width=.49\textwidth,clip=]{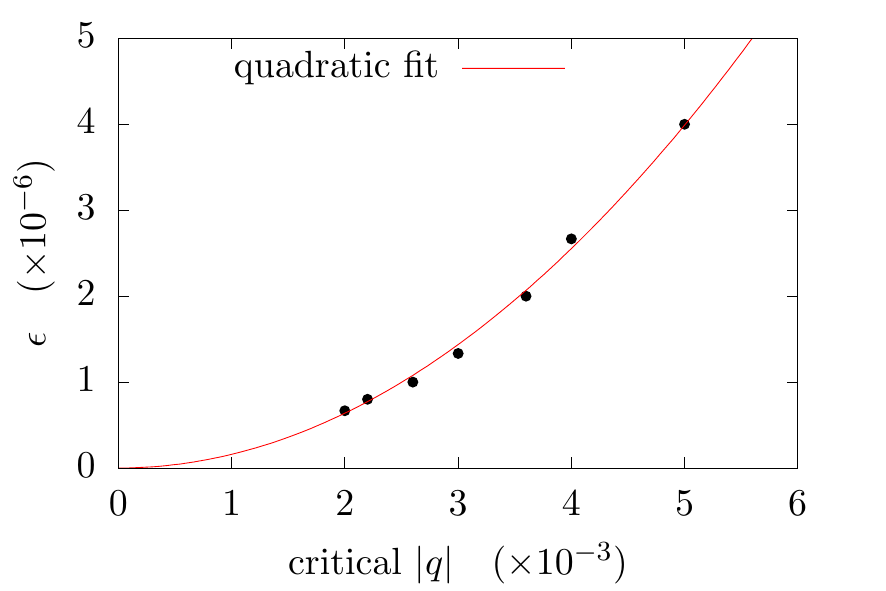}
    \caption{Top left: Oscillations of $P_\text{na}$ at large values of turns $N$ for different values of $q$. Top right: Minimum, average, and maximum value of $P_\text{na}$ for $\num{9e4}\le N \le \num{1e5}$ for different values of $q$. Center right: zoom in the small $q$ region of the average value of $P_\text{na}$ with a quadratic fit. Note that for small $|q|$ values the emittance exchange breaks down, which is indicated by $P_\text{na}$ growing towards $1$ and the $q$-value corresponding to the break down is $\epsilon$-dependent. Center left: main frequency of the oscillations of $P_\text{na}$ obtained by a refined \textsc{fft} as a function of $q$. Bottom: relation between the critical value of $|q|$, for which the break down of $P_\text{na}$ is observed, and $\epsilon$. A quadratic fit to the data is also shown, which confirms the scaling law~\eqref{eq:epsquad}. The map~\eqref{eq:mapfinal} has been used, with parameters $\omega_x = 2.602,\, \omega_{y, \rm i}=2.5, \,\omega_{y, \rm f}= 2.7$, and a set of initial conditions with $\av{I_{x, \text{i}}}=10^{-4}$, $\av{I_{y, \text{i}}}=\num{4e-4}$.}
    \label{fig:osc}
\end{figure*}

Figure~\ref{fig:osc} (top left) shows these oscillations, whose amplitude and frequency are shown in the top-right and center-left plots, respectively. The oscillations of $P_\text{na}$ are characterized by a frequency, which  has been determined by using refined techniques based on \textsc{fft}~\cite{Bartolini:292773,Bartolini:316949}, that for small $q$-values is given by $(\omega_{y, \rm f}-\omega_x)/2$. The sudden jump visible in the inset, is due to the $q$-value being too small for the $\epsilon$-value used to preserve the resonance-crossing process. This is similar to what can be observed in the bottom-right plot of the same figure.

In the center-right part of Fig.~\ref{fig:osc} a zoom  of the behavior of the average $P_\text{na}$ for small $|q|$ values is shown and the characteristic scaling with $q^2$ is clearly visible, which is linked to the neglected terms that are affecting the preservation of the invariants. Note the rapid increase of $P_\text{na}$ towards $1$ when $|q| \to 0$, which originates from the linear coupling being too small for the exchange of the invariants to take place. 

\begin{figure*}[htb]
    \centering
    \includegraphics[trim=2truemm 0truemm 5truemm 2truemm,width=0.48\textwidth,clip=]{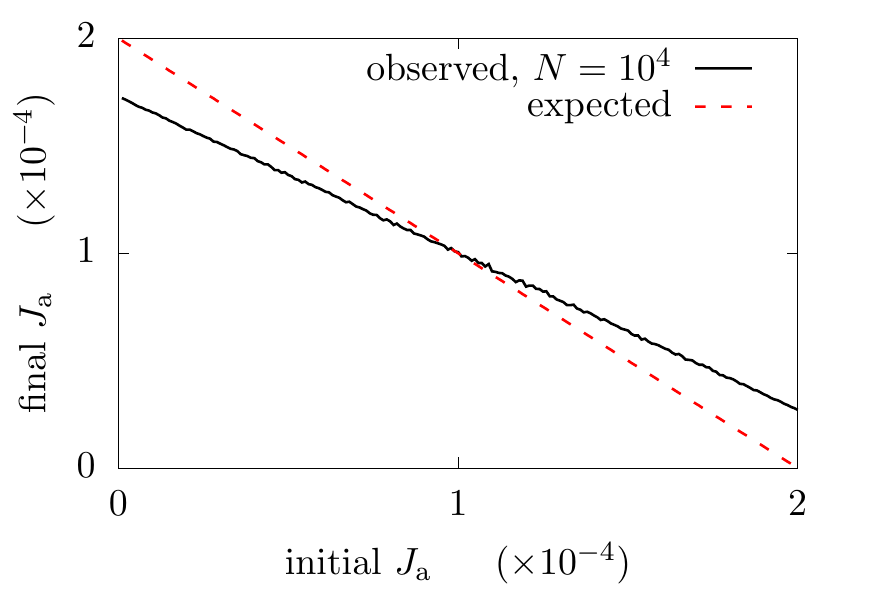}
    \includegraphics[trim=2truemm 0truemm 6truemm 2truemm,width=0.48\textwidth,clip=]{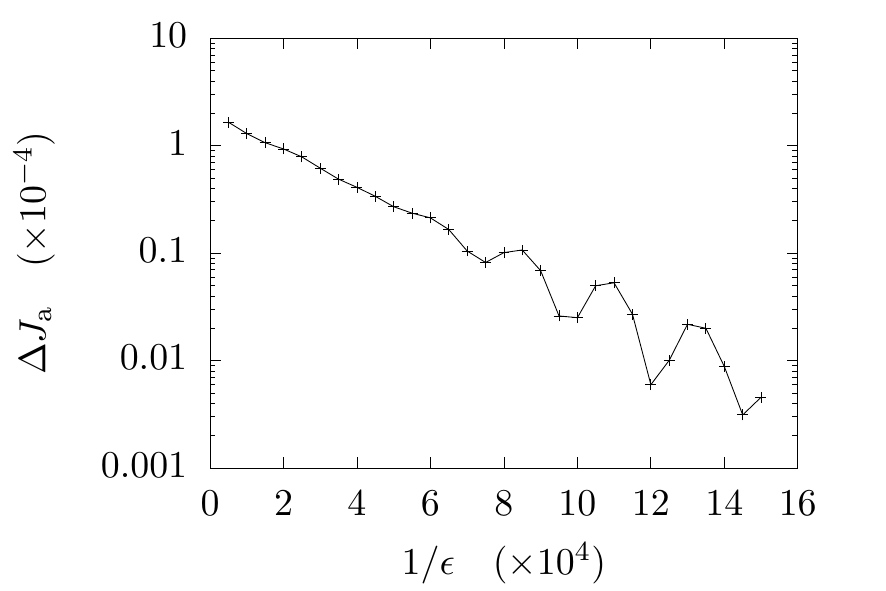}
    \caption{Left: computed, ($\av{J_{\rm a, f}}$,  black line) and expected, ($J_{\rm b}-J_{\rm a, i}$,  red, dashed) final value of the invariant as a function of $J_{\rm a, i}$ after a resonance-crossing procedure. To each $J_{\rm a, i}$ value is associated a uniform distribution of angles in $[0, 2\pi]$. Right: amplitude of the difference between the two lines in the left plot for $J_{\rm a, i}=0$ (close to the fixed-point position) as a function of $1/\epsilon$. The map~\eqref{eq:mapfinal} has been used, with parameters  $q=-0.005$, $\omega_x = 2.602,\, \omega_{y, \rm i}=2.5, \,\omega_{y, \rm f}= 2.7$, and a set of initial conditions with $J_{\rm b}=2\times 10^{-4}$.}
    \label{fig:twolines}
\end{figure*}

Another important point to stress is that these scaling laws are not connected with the features of the distribution of initial conditions, as the theoretical results clearly indicate that these properties are linked to the individual orbits of the Hamiltonian system. Indeed, this can be seen in Fig.~\ref{fig:twolines}. 

In fact, when plotting $\av{J_{\rm a, f}}$ as a function of $J_{\rm a, i}$ (left), we observe a linear difference between $(J_{\rm b}-J_{\rm a, i})$ and $\av{J_{\rm a, f}}$ when the adiabatic parameter $\epsilon$ is very small. This difference decreases exponentially when we vary the adiabaticity of the resonance-crossing process (right), \ie $\Delta J_\mathrm{a} = \kappa \exp \left [- \xi(\omega_{y,\text{f}}-\omega_{y,\text{i}})/\epsilon \right ]$. Note that also in this case oscillations appear when $\epsilon$ decreases as the actions $J_{a,b}$ are not the correct adiabatic invariant due to the value of $q$ used in the simulations.

\begin{figure*}[htb]
    \centering
    \includegraphics[trim=2truemm 0truemm 6truemm 2truemm,width=0.48\textwidth,clip=]{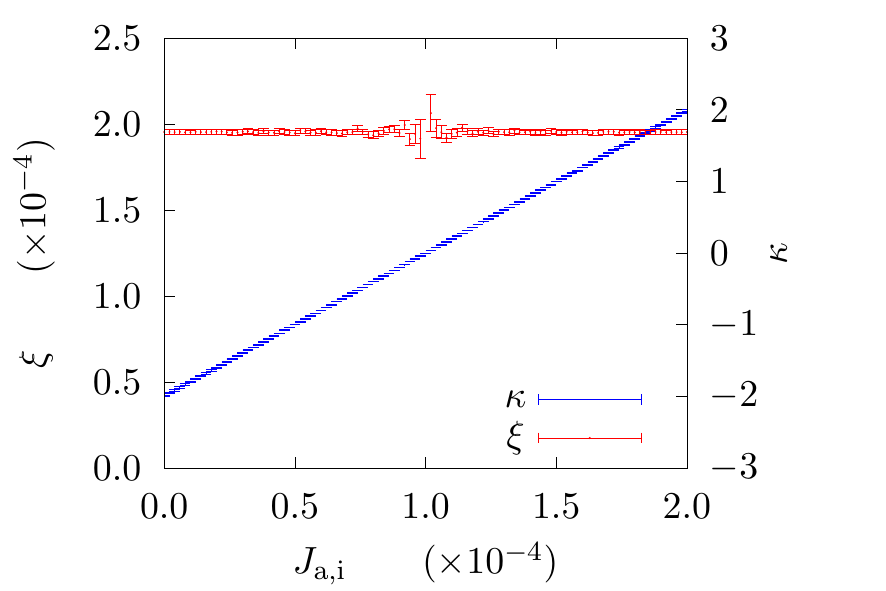}
    \includegraphics[trim=2truemm 0truemm 6truemm 2truemm,width=0.48\textwidth,clip=]{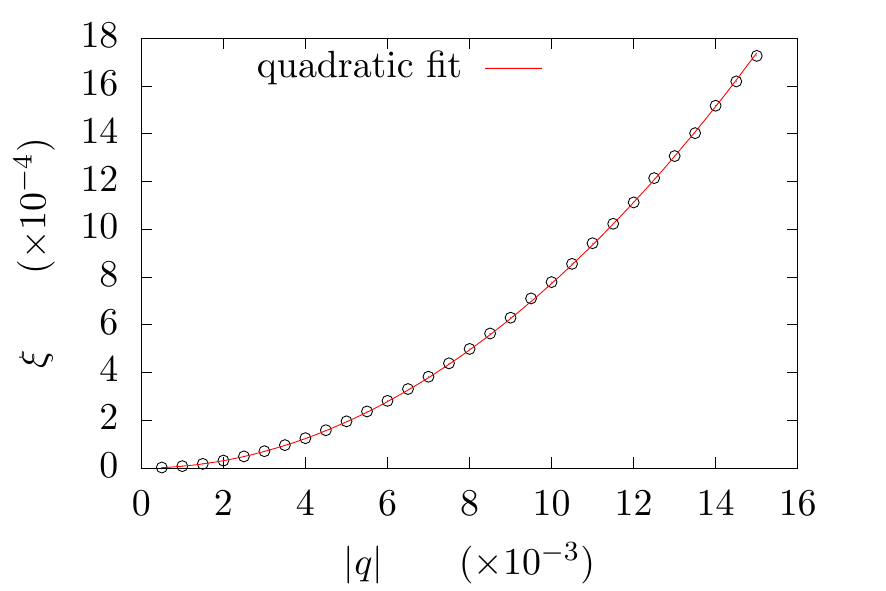}
    \caption{Left: Parameters of the exponential fit to the formula $\Delta J_\mathrm{a} = \kappa \exp \left [- \xi(\omega_{y,\text{f}}-\omega_{y,\text{i}})/\epsilon \right ]$ for distributions with constant $J_\mathrm{a, i}$ as in the bottom plot of Fig.~\ref{fig:twolines} as a function $J_\mathrm{a, i}$. The oscillations of the value of $\xi$ when $J_\text{a,i}$ is close to $\Jb/2$ is linked to the increase of the uncertainty on the exponential fit parameter at small values of $|\kappa|$. Right: Dependence of the parameter of the exponential $\xi$ on the $q$ (a quadratic fit is also presented). The map~\eqref{eq:mapfinal} has been used with parameters $q=-0.005$ (for the left plot), $\omega_x = 2.602,\, \omega_{y, \rm i}=2.5, \,\omega_{y, \rm f}= 2.7$, and a set of initial conditions with $J_{\rm b}=2\times 10^{-4}$.}
    \label{fig:abfit}
\end{figure*}

The dependence of the exponential fit parameters $\kappa$ and $\xi$ can be fully determined by means of numerical simulations. If we perform the same analysis on different values of the initial action (using a set of initial conditions $\delta(J_a-J_\mathrm{a, i})$ uniformly distributed \wrt the angular variable) and we apply the same exponential fit we see that $\kappa \propto J_\mathrm{a,i}$ whereas $\xi$ remains constant (Fig.~\ref{fig:abfit}, left). 

Therefore, the value of $\xi = \num{1.9544 \pm 0.0007e-4}$ (obtained by fitting the data with a constant) is independent on the initial radial distribution and is indeed retrieved in the fit of Fig.~\ref{fig:varnum} for $q=-0.005$ where we get $\xi = \num{1.9563 \pm 0.0015e-4}$. Finally, as predicted in Section~\ref{sec:digression}, we observe (Fig.~\ref{fig:abfit}, right) that $\xi$ depends quadratically on the linear coupling strength $q$, as found also  in~\cite{PhysRevAccelBeams.23.044003}. 

It is key to stress that the scaling law $\Delta J_\mathrm{a} = \kappa \exp \left (- \hat{\xi}\, q^2 /\epsilon \right )$, which has been justified theoretically in the previous sections, is essential in establishing a link between the two parameters $q$ and $\epsilon$ that are governing the dynamics of the system under consideration. Indeed, whenever the following is satisfied
\begin{equation}
    \epsilon = \text{const.} \times q^2
    \label{eq:epsquad}
\end{equation}
the variation of $J_\mathrm{a}$ is left unchanged. This means that when $q$ is decreased, $\epsilon$ should be reduced even further to maintain the character of the dynamics unaffected. This aspect is clearly appreciated in Fig.~\ref{fig:osc} (center right) where the curves corresponding to two values of $\epsilon$ are shown: a reduction of $\epsilon$ allows moving forward the break down behavior observed. This phenomenon has been studied in detail, and the results are shown in the bottom plot of Fig.~\ref{fig:osc}, where the relationship between the break-down value $q$ and the corresponding $\epsilon$ value is shown. The points lie on a quadratic curve, in perfect agreement with the scaling law reported in Eq.~\eqref{eq:epsquad}. It is evident that in the case of a nonlinear crossing of the resonance, the relationship~\eqref{eq:epsquad} reads
\begin{equation}
    \epsilon = \text{const.} \times q^\frac{2n+2}{2n+1} \, ,
    \label{eq:epspoly}
\end{equation}
where $2n+1$ is the power of $\lambda$ with which the resonance is crossed.

\begin{figure}[htb]
    \centering
    \includegraphics[trim=0truemm 0truemm 6truemm 2truemm,width=0.7\textwidth,clip=]{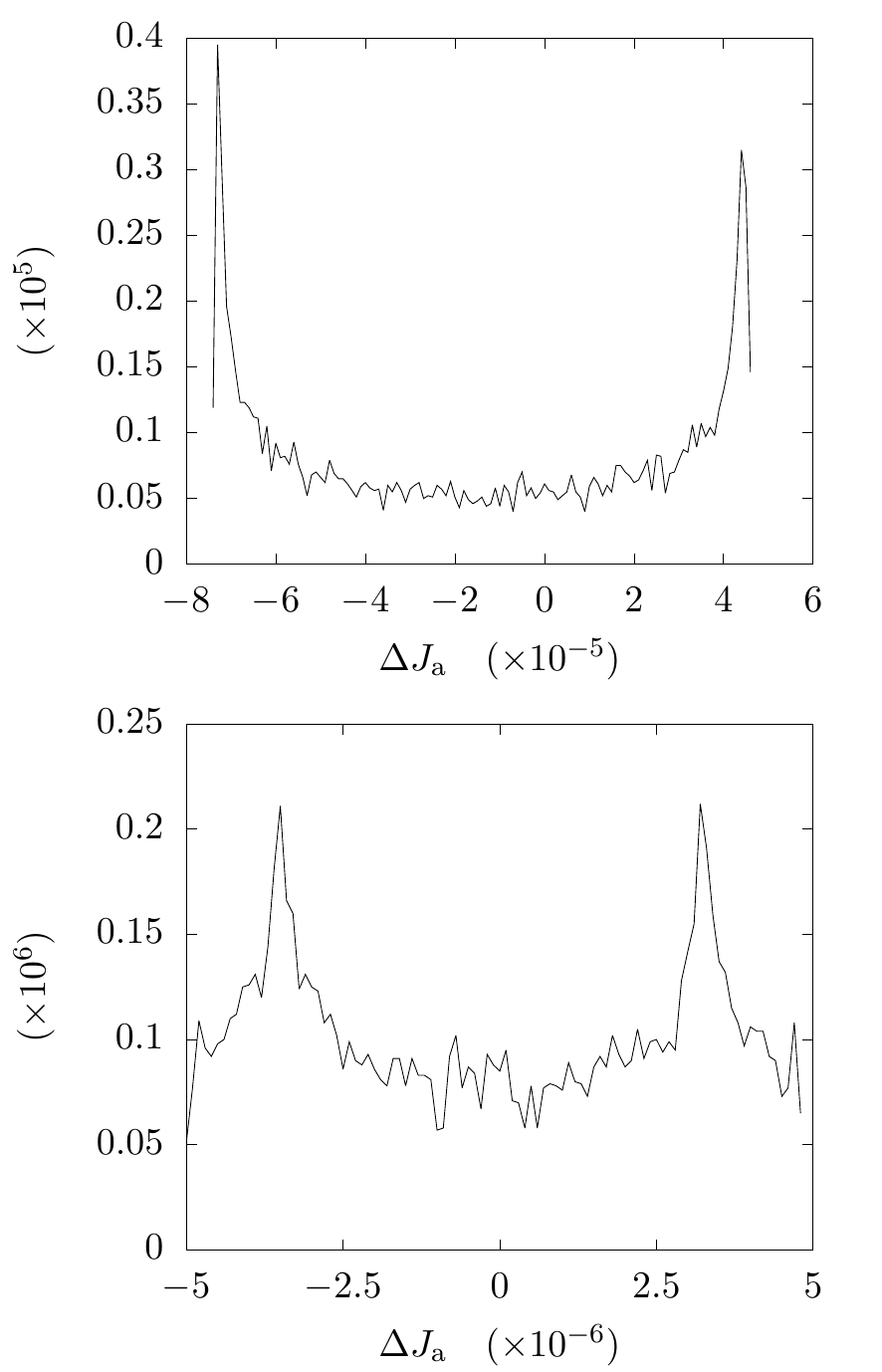}
    \caption{Distribution of the jump of the invariant (difference between $J_\mathrm{a, f}$ and the expected one $J_\mathrm{b}-J_\mathrm{a, i}$) for particles at $J_{1, \rm i}=5\times 10^{-5}$. Top: $N = 10^{4}$, \ie $\epsilon = \num{2e-5}$, Bottom: $N=5\times 10^{4}$, \ie $\epsilon=\num{4e-6}$. The map~\eqref{eq:mapfinal} has been used, with parameters $q=-0.005$, $\omega_x = 2.602,\, \omega_{y, \rm i}=2.5, \,\omega_{y, \rm f}= 2.7$, $J_{\rm b}=2\times 10^{-4}$.} 
    \label{fig:distr}
\end{figure}

The distribution of the action jumps during the resonance-crossing process for a set of initial conditions with the same value of $J_\mathrm{a, i}$ and the phase uniformly distributed in $[0, 2\pi]$, is shown in Fig.~\ref{fig:distr}. 

It is worth noting that when $\epsilon$ decreases the average of the jump of the invariant tends to zero and the extent of the support of the distribution of $\Delta J_\mathrm{a}$ shrinks. At the same time, the distribution becomes more symmetrical and it flattens out. 
\subsection{Nonlinear map model}\label{sec:detuning_simulations}

As described in Section~\ref{sec:detuning}, we studied also the impact of detuning with amplitude on the adiabaticity of the emittance exchange using the map model given in~\eqref{eq:mapfinal_oct}, introducing a normal octupole with normalized strength $k_3=K_3\beta_x^2/6$, setting $\chi=1$, and simulating the resonance-crossing process in the same way as in absence of detuning.

\begin{figure}[htb]
    \centering
    \includegraphics[trim=-5truemm 0truemm 6truemm 2truemm,width=.49\textwidth,clip=]{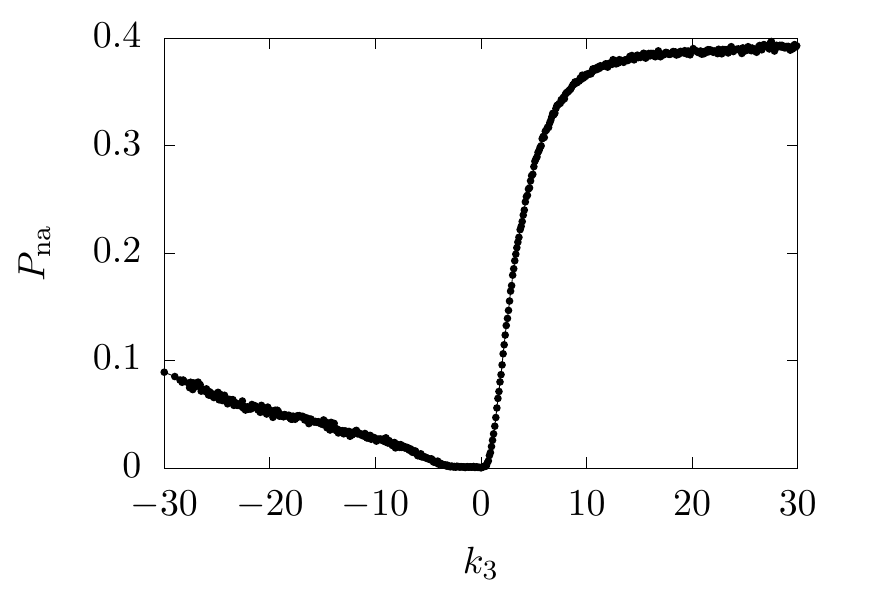}
    \includegraphics[trim=1truemm 0truemm 6truemm 2truemm,width=.49\textwidth,clip=]{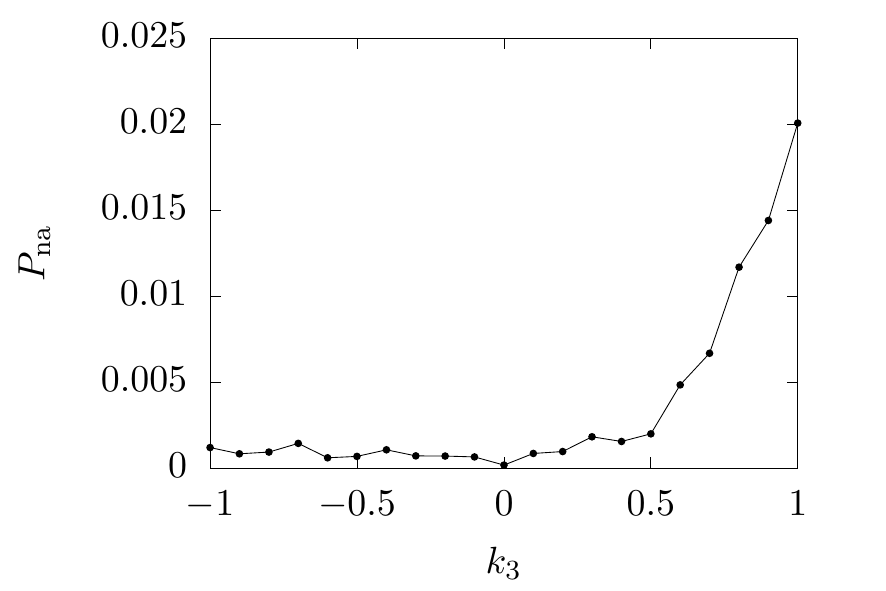}
    \caption{Top: $P_\text{na}$ as a function of the octupole strength coefficient $k_3$. Bottom: zoom in of the top plot in the range $-1\le k_3 \le 1$. The map~\eqref{eq:mapfinal_oct} has been used with parameters $q=-0.005$, $\omega_x = 2.602,\, \omega_{y, \rm i}=2.5, \,\omega_{y, \rm f}= 2.7$, $N=10^5$, and a set of initial conditions with $\av{I_{x, \rm i}}=10^{-4}$, $\av{I_{y, \rm i}}=4\times 10^{-4}$.}
    \label{fig:oct_pna}
\end{figure}

For Gaussian distributions of initial conditions  corresponding to different emittances in $x$ and $y$, $P_\text{na}$ has been evaluated for different values of $k_3$ and the results are shown in Fig.~\ref{fig:oct_pna} (top), whereas a zoom in is provided in the bottom plot. 

Two behaviors, depending on the value and sign of $k_3$, are clearly visible. For $k_3<0$ and $k_3$ small in absolute value, $P_\text{na}$ is essentially zero, which indicates that even when nonlinear effects are present, a perfect emittance exchange occurs. Such a behavior is hardly seen for $k_3>0$, even in the neighborhood of zero. Rather, a sharp rise of $P_\text{na}$ is visible. Globally, outside a small interval around zero for $k_3$, $P_\text{na}$ is always different from zero, indicating that the emittance exchange is not perfect. 

Note that the position of the new fixed points either on the right or on the left of the coupling arc is linked to the sign of $k_3$. Therefore, the presence of the new detuning-related fixed points has different impact on the final distribution, and hence on the emittance exchange, depending on where they are located in phase space. In this specific case, when $0<k_3<0.5$ even if $k_3$ is positive the effect on $P_\text{na}$ is negligible: this is compatible with the theoretical predictions that small enough values of $k_3$ do not generate new fixed points and do not affect the emittance exchange.

\begin{figure*}
\centering
    \includegraphics[trim=1truemm 0truemm 6truemm 2truemm,width=.47\textwidth,clip=]{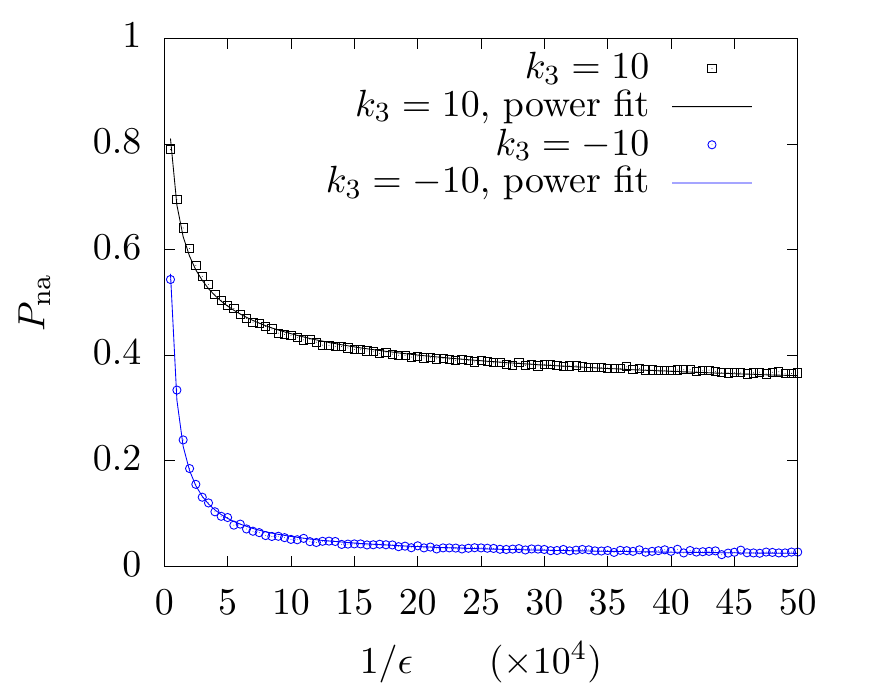}
    \includegraphics[trim=1truemm 0truemm 2truemm 2truemm,width=.47\textwidth,clip=]{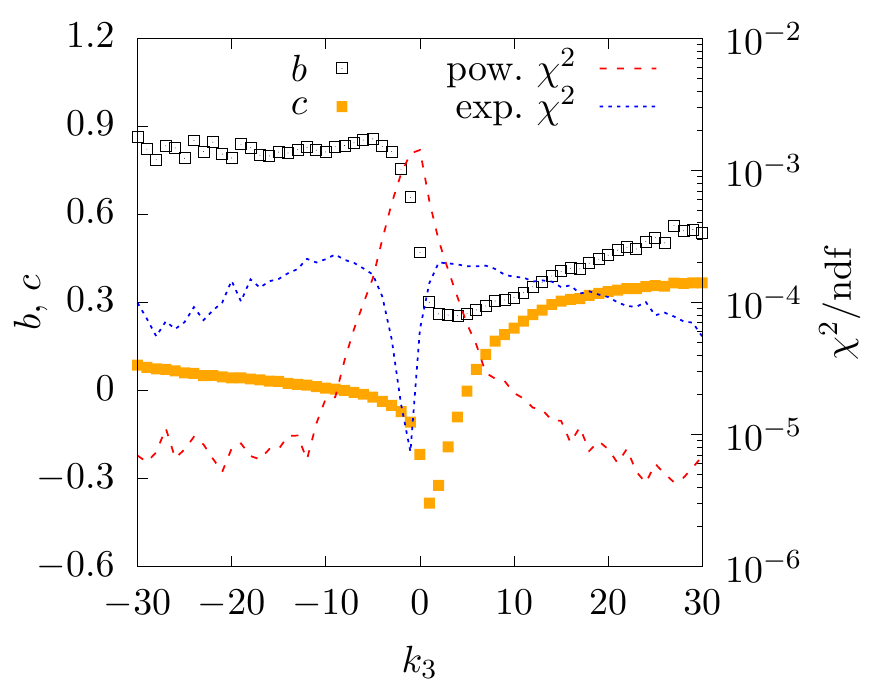}\\
    \includegraphics[trim=1truemm 0truemm 6truemm 2truemm,width=.49\textwidth,clip=]{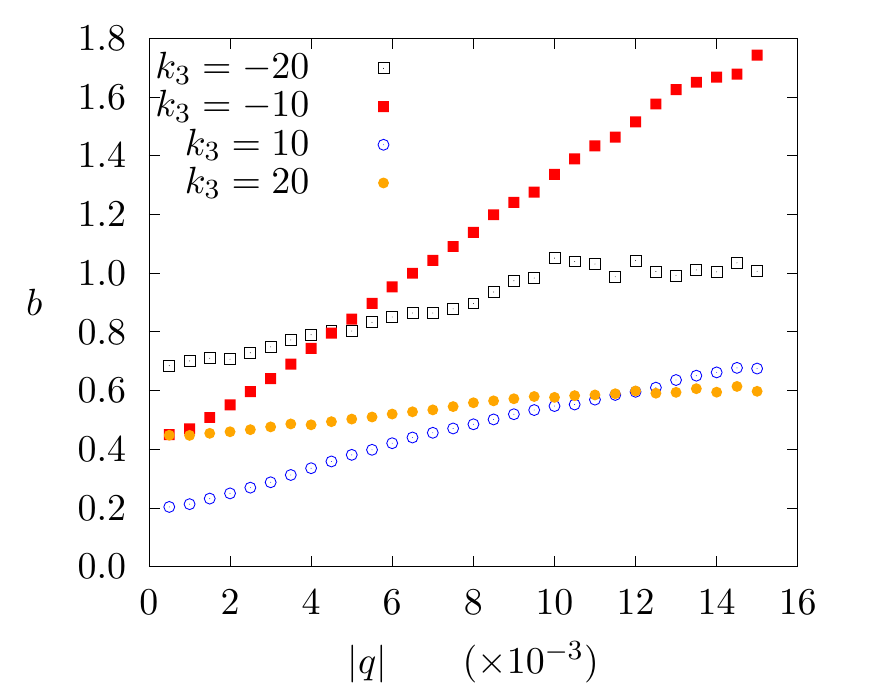}
    \includegraphics[trim=1truemm 0truemm 6truemm 2truemm,width=.49\textwidth,clip=]{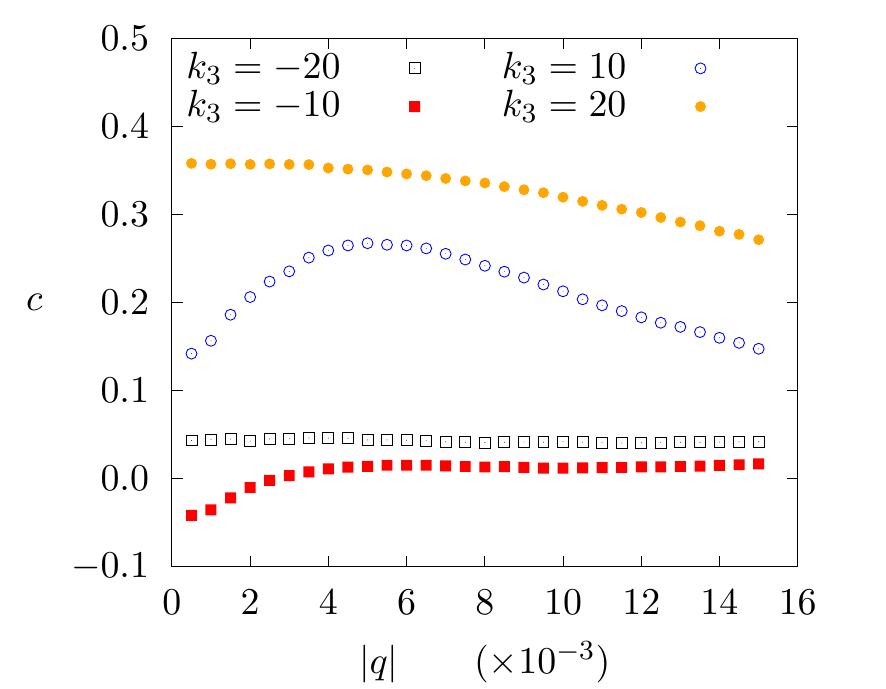}
    \caption{Top left: Dependence of $P_\text{na}$ on $\epsilon$ for $k_3=10$ and $k_3=-10$. A power-law dependence $P_\text{na}=a\epsilon^b + c$, is fitted and the results also shown ($b=\num{-0.381\pm0.009} $ for $k_3=10$ and $b=\num{-0.844\pm0.009}$ for $k_3=-10$). Top right: Dependence of the power-law exponent $b$ and $c$ (left axis) and reduced $\chi^2$ for the power- and exponential-law fits (right axis) as a function of $k_3$.
    Bottom: dependence of power-law exponent $b$ (left) and $c$ (right) on $q$, for some values of $k_3$. The map~\eqref{eq:mapfinal_oct} has been used, with parameters $q=-0.005$ (except for the bottom plots), $\omega_x = 2.602,\, \omega_{y, \rm i}=2.5, \,\omega_{y, \rm f}= 2.7$, $N=10^5$, $\epsilon=\num{2e-6}$, and a set of initial conditions with $\av{I_{x, \rm i}}=10^{-4}$, $\av{I_{y, \rm i}}=4\times 10^{-4}$.}
    \label{fig:detuning_powfit}
\end{figure*}

The essential difference between the linear and nonlinear cases is clearly visible when investigating the dependence of $P_\text{na}$ on the adiabatic parameter $\epsilon$. This is shown in Fig.~\ref{fig:detuning_powfit} (top left). The exponential behavior is lost and is replaced by a power-law function for $P_\text{na}$. 

The two values of $k_3$ have been selected to provide cases in which more than two fixed points are present. The dependence on $|k_3|$ is shown in the top-right plot, where the dependence of the fit parameters is shown together with the behavior of the reduced $\chi^2$ is also reported for the power- and exponential-law cases. The behavior of the reduced $\chi^2$ for the two fit models shows clearly that while the exponential dependence is the most suitable one for $k_3$ close to zero, the power-law best describes the data outside this interval of $k_3$. This is in full agreement with the theoretical discussion carried out in the previous section. Finally, the dependence is shown as a function of $q$ in the bottom plots of Fig.~\ref{fig:detuning_powfit} for some values of $k_3$.

Also in presence of nonlinear detuning with amplitude, it is possible to devise a scaling law linking $q$ and $\epsilon$. It is very easy to conclude that $P_\text{na}$ is linked to the change of the invariants during the crossing process. Therefore, the scaling law $P_\text{na}=a\epsilon^{b(k_3,q)} + c(k_3,q)$, which has been analyzed in Fig.~\ref{fig:detuning_powfit}, gives rise to the following relationship
\begin{equation}
    \ln \epsilon = \frac{\text{const.}+ c(k_3,q)}{a \, b(k_3,q)} \, , 
\end{equation}
which should be fulfilled in order to keep constant the change of the invariant. The essential difference with respect to what has been found in the absence of nonlinear detuning is apparent.

\begin{figure*}[htb]
\centering
    \includegraphics[trim=1truemm 0truemm 6truemm 2truemm,width=.49\textwidth,clip=]{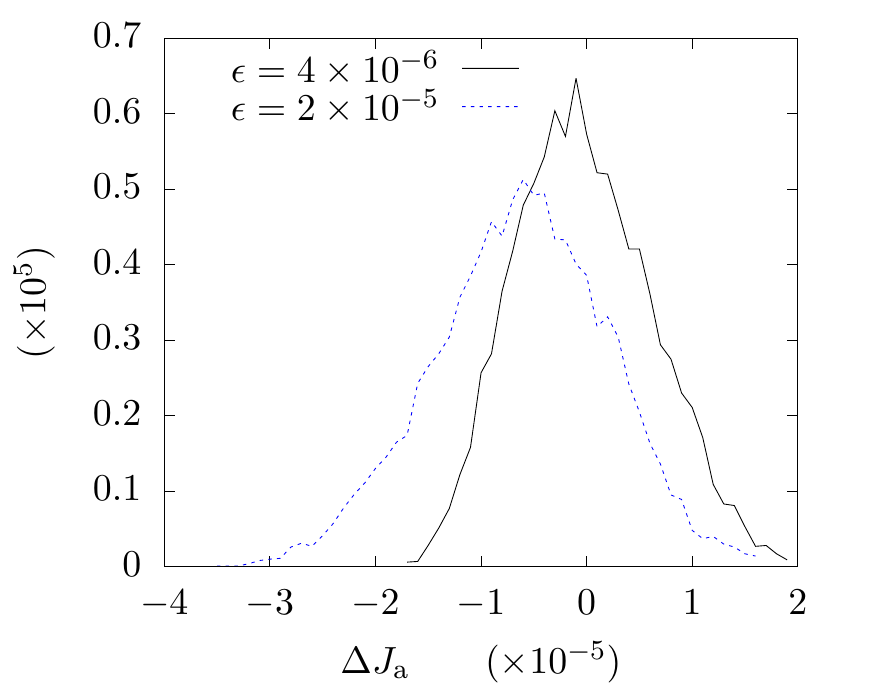}
    \includegraphics[trim=1truemm 0truemm 6truemm 2truemm,width=.49\textwidth,clip=]{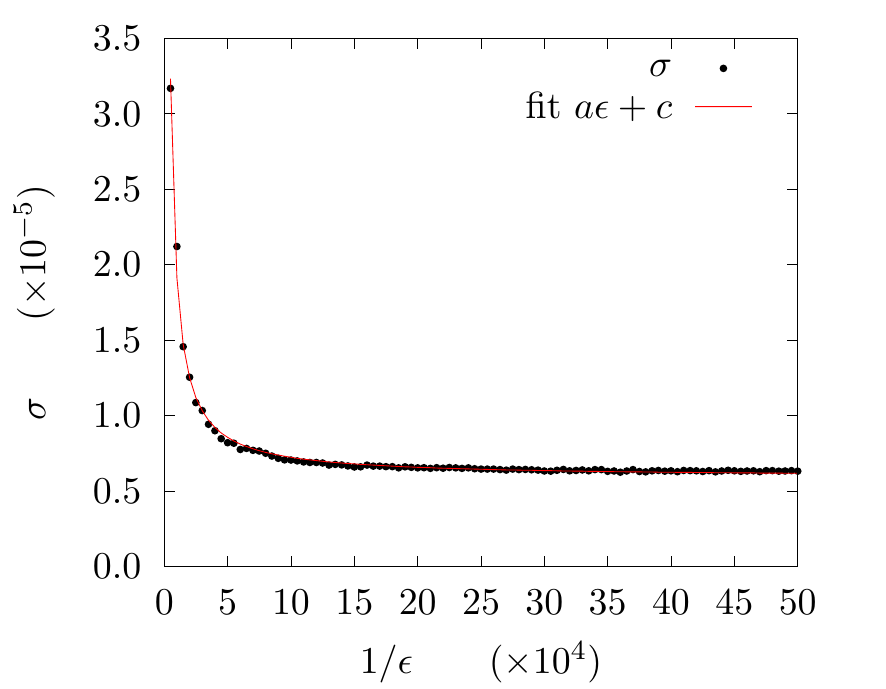}
    \includegraphics[trim=1truemm 0truemm 4truemm 0truemm,width=.49\textwidth,clip=]{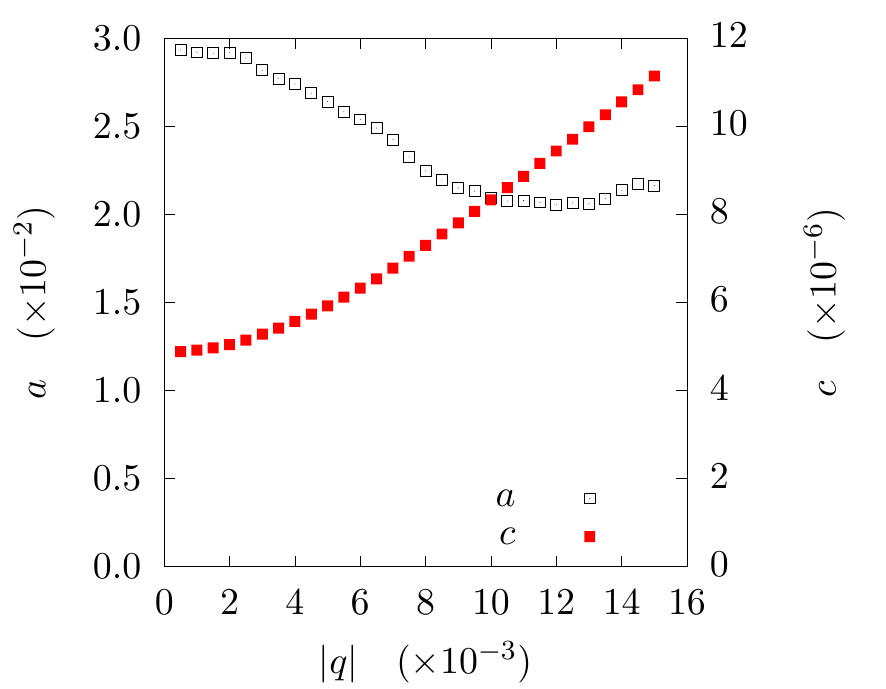}
    \caption{Top left: Distribution of the the jump of the invariant (difference between $J_\mathrm{a, f}$ and the expected one $J_\mathrm{b}-J_\mathrm{a, i}$) for $k_3=100$, for $N=5\times 10^4$, \ie $\epsilon = \num{4e-6}$ and $N=10^4$, \ie $\epsilon = \num{2e-5}$. Top right: Standard deviation $\sigma$ of the jumps distribution for $k_3=100$ at different values of $\epsilon$. A inverse linear fit is presented. Bottom: dependence of the fit parameters $a$ and $c$ on $q$, for $k_3=100$ and using the same initial conditions of the previous plots. The map~\eqref{eq:mapfinal_oct} has been used, with parameters $q=-0.005$ (except for the bottom plot), $\omega_x = 2.602,\, \omega_{y, \rm i}=2.5, \,\omega_{y, \rm f}= 2.7$, and a set of initial conditions with $J_\mathrm{a, i}=5\times 10^{-5}$, $J_\mathrm{b}=2\times 10^{-4}$.}
    \label{fig:oct_distrib}
\end{figure*}

Finally, Fig.~\ref{fig:oct_distrib} shows the features of the distribution of the jumps of the invariants in the nonlinear case. 

In the top-left plot, the distribution of $J_\mathrm{a}$, \ie the difference between the observed and the expected value of $J_\mathrm{a,f}$, for $k_3\gg 1$ for initial conditions at a given fixed $J_\mathrm{a,i}$ and $J_\mathrm{b}$ is shown for two values of $\epsilon$. The distributions look completely different from the ones obtained in the linear case (see Fig.~\ref{fig:distr}). In fact, although the distributions tend to be centered around zero as $\epsilon$ decreases, their standard deviation $\sigma$ as a function of $\epsilon$ tends to a finite limit for $\epsilon \to 0$. This indicates that the invariant is indeed experiencing a finite jump even when the adiabatic regime is reached. This is a direct consequence of the presence of a separatrix linked with the additional fixed points, as found in \cite{NEISHTADT1986,ITIN2007108}. The scaling law of $\sigma$ is represented by a power-law dependence on $\epsilon$, as shown in the top-right plot, whose parameters are functions of $q$, according to the functional behavior shown in the bottom plot of Fig.~\ref{fig:oct_distrib}. We remark that such a behavior is model dependent.
\section{Conclusions} \label{sec:conc}
In this paper, the Hamiltonian theory of the dynamic crossing of the coupling resonance has been presented and discussed in detail, considering not only the linear, but also the nonlinear case. 

The main focus has been the analysis of the so-called emittance exchange process, which arises from the resonance crossing. The detail of the mechanism has been considered, both in standard phase space as well as considering the normal modes, which provide the correct description of the system behavior. In particular, this framework allows considering the interplay between the two small parameters of the problem, namely the adiabaticity parameter and the strength of the linear coupling. It is worth stressing that, while the true invariants are exactly exchanged, the linear actions feature only a partial exchange. This is because they are not the true invariants of the system and that a projection effect from the space of the invariants to the physical space has to be taken into account. All this should be carefully considered when translating these observations to  circular accelerators, in which the physical planes are normally the reference concepts used to interpret the phenomena linked with the crossing of the  coupling resonance. Otherwise, the system undergoes periodic variations of the transverse emittances as a function of $\epsilon^{-1}$. This is the simple consequence of the fact that, whenever the coupling strength is not small, the uncoupled emittances are no longer the correct invariants of the system. Furthermore, it has been shown how the dynamical properties of the system under consideration can be best appreciated by looking at the dynamics on a sphere, rather than the standard flat phase space. 

It has been discussed how the presence of a real separatrix in phase space is the key feature that distinguishes the behavior of the crossing of the coupling resonance for a linear and nonlinear system. The origin of such a difference is connected with the breaking down of the analytical properties of the dynamical system whenever a separatrix is present.

Detailed numerical simulations have been carried out to provide a full characterization of the rich spectrum of behaviors. It is worth stressing that the observed exponential dependence on the adiabaticity parameter of the emittance exchange is a natural consequence only of the analyticity properties of the system under consideration. To the best of our knowledge, for the first time, the behavior of a nonlinear system has been probed while crossing the coupling resonance and, in excellent agreement with the theory presented, a power law, instead of an exponential one, has been observed for the emittance exchange process. 

A fundamental relationship between the key system parameters $q$ and $\epsilon$ has been derived for the case with and without nonlinear detuning. Such a link describes how the crossing of the coupling resonance should be performed in order to keep it adiabatic as a function of the value of the linear coupling. It is evident that such a relationship is of paramount importance in applications and, to the best of our knowledge, it has been derived for the first time in this paper.

Finally, a digression has been made, considering the features of a two-way crossing of the coupling resonance. By applying the adiabatic theory it has been shown that reversibility of the resonance crossing process is not granted in the nonlinear case, even in the adiabatic limit. This is a rather interesting and thoughtful result for its implications in the domain of accelerator physics applications, \eg in the case of periodic resonance crossing induced by finite chromaticity, for which a non-negligible impact on the beam distribution is to be expected, no matter the speed of the resonance crossing.
\clearpage
\appendix
\section{Comments on the equations of motion of the hamiltonian~\texorpdfstring{\eqref{eq:ham_fundamental}}{}} \label{sec:app1}
Let us consider a Hamiltonian $H(\phi, J)$ and the scaled Hamiltonian $\tilde H(\phi, J)=\lambda(J)(H(\phi, J)-E_0)$, where $H(\phi_0, J_0)=E_0$. We explicitly compute (in this context $'$ stands for the derivative with respect to $J$)
\begin{equation}
\begin{split}
\frac{\partial \tilde H}{\partial J} & =\lambda(J)\frac{\partial H}{\partial J}+\lambda'(J)(H(\phi, J)-E_0) \nonumber \\
\\
\frac{\partial \tilde H}{\partial \phi} & =\lambda(J)\frac{\partial H}{\partial \phi} \nonumber
\end{split}
\end{equation}
so that given the initial conditions $(\phi_0, J_0)$  and $E_0$ the solution of the Hamiltonian system
\begin{equation}
\begin{split}
\frac{\dd \phi}{\dd \tau} & =\phantom{-}\frac{\partial \tilde H}{\partial J}\nonumber \\
\\
\frac{\dd J}{\dd \tau} & =-\frac{\partial \tilde H}{\partial \phi}\nonumber 
\end{split}
\end{equation}
with initial energy $\tilde H(\phi, J) = 0$, we obtain the system 
\begin{equation}
\begin{split}
\frac{1}{\lambda(J)}\frac{\dd \phi}{\dd \tau} & = \phantom{-} \frac{\partial H}{\partial J}\nonumber \\
\frac{1}{\lambda(J)}\frac{\dd J}{\dd \tau} & =-\frac{\partial H}{\partial \phi}\nonumber 
\end{split}
\end{equation}
and if we introduce the scaled time $\lambda(J)\dd\tau=\dd t$, we recover the solution of the initial Hamiltonian. Of course, the phase-space structure of both Hamiltonian systems is the same. 

Using this approach, the solutions of the equations of motion of the Hamiltonian~\eqref{eq:ham_fundamental} can be associated to those of the Hamiltonian
\begin{equation}
\begin{split}
\tilde H(\phi, J,\lambda) &=\frac{1}{\sqrt{(1-J)J}}\left \{ \left [\delta(\lambda) J-E_0 \right ] + \right . \\
& +  \left . q \sqrt{(1-J)J}\sin\phi\right \} \\
&=\frac{\delta J-E_0}{\sqrt{(1-J)J}}+q \sin\phi
\end{split}
\end{equation}
in the scaled time $\sqrt{(1-J)J}\dd t=\dd\tau$, where $E_0$ is the value of the initial energy, namely
\begin{equation}
E_0=\delta(\lambda)J_0+ q \sqrt{(1-J_0)J_0}\cos\phi_0 \, .
\label{eq:ham_e0}
\end{equation}
Therefore we consider the orbit with zero energy $H'=0$ and the equations of motion give the solution of the initial Hamiltonian system
in the scaled time. But the phase space has the same structure for both Hamiltonians. In the Hamiltonian $H'$, $E_0$ is a parameter.

Starting from the Hamiltonian \eqref{eq:ham_fundamental} and using the expression given in Eq.~\eqref{eq:ham_e0} for $E_0$  we get the equation of motion (see \cite{Lee:2651939})
\begin{equation} 
\ddot J + (\delta^2+q^2) J = \delta E_0 + \frac{q^2}{2} 
\end{equation}
that can be explicitly solved in the form of
\begin{equation}
\begin{split}
J(t) & = \sqrt{ \qty(\delta E_0 + \frac{q^2}{2})-\frac{E_0^2}{\delta^2+q^2}} \cos(\sqrt{\delta^2+q^2}t) + \\
+ & \delta E_0 + \frac{q^2}{2}
\end{split}
\end{equation}
that is a sinusoidal function, whose period is given by
\begin{equation}
T = \frac{2\pi}{\sqrt{\delta^2+q^2}}
\end{equation}
and that is independent on the initial energy.
\section{Detail of the analysis of the dynamics using the normal modes} \label{sec:app2}
Let us start from the Hamiltonian~\eqref{eq:hamq} with the goal of considering the dynamics considering the normal modes. Let us denote $\bm{\omega} = (\omega_1,\,\omega_2)$ the vector of the normal modes and $\bm{R}(\lambda)$ the orthogonal matrix built using the components of the eigenvectors, then the normal variables $\bm X$ are defined
\begin{equation}
\bm x=\bm R(\lambda) \, \bm X
\end{equation}
A generating function for the transformation $F_2( \bm x,\bm P, \lambda)$ can be written in the form
\begin{equation}
F_2(\bm x,\bm P, \lambda)=\bm P \bm R(\lambda)^\top \bm x 
\end{equation}
and the new Hamiltonian reads
\begin{equation}
\begin{split}
H(\bm X,\bm P,\lambda) &=\frac{{P_1^2+P_2^2}}{2}+\frac{\omega_1^2(\lambda)X_1^2 + \omega_2^2(\lambda)X_2^2}{2} + \\
&+\epsilon \bm P \pdv{\bm{R}^\top}{\lambda} \bm R \, ,
\end{split}
\end{equation}

where the last term is generated by the time derivative of the generating function. Furthermore, it can be verified that the matrix 
\begin{equation}
\bm \Xi(\lambda)=\pdv{\bm{R}^\top}{\lambda}\bm{R}
\end{equation}
is anti-symmetric and has the form
\begin{equation}
\bm\Xi(\lambda)=\begin{pmatrix}0 & \xi(\lambda) \\ -\xi(\lambda) & 0\end{pmatrix} \, ,
\end{equation}
where 
\begin{equation}
\begin{split}
    \xi(\lambda)&=-2q\frac{\omega_x(\lambda)\,\omega_x'(\lambda)-\omega_y(\lambda)\,\omega_y'(\lambda)}{\left (\omega_x^2(\lambda)-\omega_y^2(\lambda) \right )^2+4q^2} \\
&=-q\frac{\delta_2'(\lambda)}{\delta_2^2(\lambda)+4q^2}
\end{split}
\end{equation}
and we obtain a term analogous to the Coriolis potential in the Hamiltonian. Note that $\xi(\lambda)=0$ when $\delta'(\lambda)=0$ and $\xi(\lambda^\ast)\propto -(2q)^{-1}\gg 1$ when $\delta(\lambda^\ast)=0$, however, in this case $\delta'(\lambda^\ast)=O(1)$.

We evaluate
\begin{equation}
\frac{\omega_1}{\omega_2}=\frac{1}{2}\left ( \frac{\omega_1^2+\omega_2^2}{\omega_1\omega_2}\right )\left [1+
\frac{\omega_1^2-\omega_2^2}{\omega_1^2+\omega_2^2}\right ]
\end{equation}
so that we get the estimates
\begin{equation}
\begin{split}
\sqrt{\frac{\omega_1}{\omega_2}}&\simeq\sqrt{\frac{1}{2}\left ( \frac{\omega_1^2+\omega_2^2}{\omega_1\omega_2}\right )}
\left [1+\frac{1}{2}\frac{\omega_1^2-\omega_2^2}{\omega_1^2+\omega_2^2}\right ] \\
&=\sqrt{\frac{1}{2}\left ( \frac{\omega_x^2+\omega_y^2}{\sqrt{\omega_x^2\omega_y^2-q^2}}\right )}
\left [1+\frac{1}{2}\frac{\sqrt{\delta_2^2(\lambda)+4q^2}}{\omega_x^2+\omega_y^2}\right ] \\
\sqrt{\frac{\omega_2}{\omega_1}}& \simeq \sqrt{\frac{1}{2}\left ( \frac{\omega_x^2+\omega_y^2}{\sqrt{\omega_x^2\omega_y^2-q^2}}\right )}
\left [1-\frac{1}{2}\frac{\sqrt{\hat\delta^2(\lambda)+4q^2}}{\omega_x^2+\omega_y^2}\right ] \, .
\end{split}
\end{equation}
Finally we define the quantities
\begin{eqnarray}
\xi_1(\lambda)&=&\xi(\lambda)\sqrt{\frac{1}{2}\left ( \frac{\omega_x^2+\omega_y^2}{\sqrt{\omega_x^2\omega_y^2-q^2}}\right )}
 \\
\xi_2(\lambda)&=&\xi(\lambda)\sqrt{\frac{1}{8}\frac{\delta^2
_2(\lambda)+4q^2}
{\sqrt{(\omega_x^2\omega_y^2-q^2})(\omega_x^2+\omega_y^2)}} \, ,
\end{eqnarray}
and we observe that $\xi_2=O(q)$ when $\delta(\lambda)\to 0$.

By introducing the scaled variables $\bm{\tilde X} = (X_1/\sqrt \omega_1,\,X_2/\sqrt \omega_2)$, $\bm{\tilde P} = (P_1\,\sqrt \omega_1,\,P_2\,\sqrt \omega_2)$ and the matrix $\bm\Omega = \mathrm{diag}(\omega_1,\,\omega_2)$ we get the Hamiltonian
\begin{equation}
\begin{split}
H(\bm {\tilde X},\bm {\tilde P},\lambda) &=\omega_1(\lambda)\frac{X_1^2+P_1^2}{2}+\omega_2(\lambda) \frac{X_2^2+P_2^2}{2}+\\
&+ \epsilon \bm{\tilde P}^\top {\bm \Omega}\,\bm\Xi\, {\bm\Omega^{-1}}\bm{\tilde X}+ \epsilon \bm{\tilde X}^\top \dv{\bm\Omega}{\lambda}{\bm\Omega^{-1}} \bm{\tilde P} \, ,
\end{split}
\label{eq:ham_fundamental2}
\end{equation}
where the last term is generated by the time derivative of the generating function of the co-ordinate transformation and
\begin{equation}
\dv{\bm\Omega}{\lambda}{\bm\Omega^{-1}} = \begin{pmatrix} \omega'_1(\lambda)/\omega_1(\lambda) & 0 \\ 0 & \omega'_2(\lambda)/\omega_2(\lambda) \end{pmatrix} \, .
\end{equation}

The Hamiltonian~\eqref{eq:ham_fundamental2} can be cast in the following form by using the linear action-angle variables
\begin{equation}
\begin{split}
H(\bm \theta,& \bm I,\lambda) =\omega_1(\lambda)I_1+\omega_2(\lambda)I_2+2\epsilon\xi\sqrt{I_1I_2} \times \\
& \times \left(\sqrt\frac{\omega_2}{ \omega_1}\sin\theta_1\cos\theta_2-\sqrt\frac{\omega_1}{\omega_2}
\cos\theta_1\sin\theta_2\right )+\\
&+\epsilon\frac{\omega'_1}{ \omega_1}I_1\sin\theta_1\cos\theta_1+\epsilon\frac{\omega'_2}{ \omega_2}I_2
\sin\theta_2\cos\theta_2 \, .
\end{split}
\end{equation}

Both actions $I_{1,2}$ are adiabatic invariants as the resonant conditions $\omega_1\pm\omega_2=0$ are never satisfied, and we define
\begin{equation}
\begin{split}
    \hat \delta(\lambda) & =\omega_1(\lambda)-\omega_2(\lambda)\\
    & = \delta(\lambda) \frac{\omega_x(\lambda)+\omega_y(\lambda)}{\omega_1(\lambda)+\omega_2(\lambda)} -\frac{1}{\omega_1(\lambda)+\omega_2(\lambda)} \times \\
    & \times \left (\delta_2(\lambda)+\sqrt{\delta^2_2(\lambda)+4q^2} \right )
\end{split}
\end{equation}
so that $\hat \delta(\lambda) \geq O(q)$ and there is a quasi-resonant condition only if $q \ll 1$. Hence, we introduce the slow angle $\phi_1=\theta_1-\theta_2$, and we define 
$\phi_2=\theta_2$ and $J_1, J_2$ are the corresponding actions, so that the new Hamiltonian is
\begin{equation}
\begin{split}
H&(\bm \phi,\bm J,\lambda)=\hat \delta(\lambda)J_1+\omega_2(\lambda)J_2+\\
& + 2\, \epsilon \,  \sqrt{(J_2-J_1)J_1} \left [ \xi_1(\lambda) \sin\phi_1+\xi_2(\lambda) \sin(\phi_1+2\phi_2) \right ] + \\
&+ \frac{\epsilon}{2}\left [\frac{\omega'_1}{\omega_2}J_1 \sin 2(\phi_1+\phi_2)+\frac{\omega'_2}{\omega_2}(J_2-J_1)\sin 2\phi_2 \right ] \, .
\end{split}
\end{equation}

If $\omega_2\sim 1$ one can average on the angle $\phi_2$ and the Hamiltonian reduces to 
\begin{equation}
\begin{split}
H(\bm \phi,\bm J, \lambda)& =\hat \delta(\lambda)J_1+\omega_2(\lambda)J_2+ \\
& + 2\, \epsilon \, \xi_1(\lambda)\sqrt{(J_2-J_1)J_1}\sin\phi_1  \, .
\end{split}
\end{equation}
We remark that $\xi_1(\lambda)=O(q^{-1})$ and $\hat \delta(\lambda)=O(q)$ when $\delta_2(\lambda)\to 0$. Therefore, using a time scaling, the previous Hamiltonian is equivalent to 
\begin{equation}
    H(\bm \phi,\bm J,\lambda)=\gamma(\lambda)J_1+\epsilon\sqrt{(J_2-J_1)J_1}\sin\phi_1 \, ,
    \label{eq:hamsphe1}
\end{equation}
where $J_2$ is an integral of motion whereas $J_1$ changes when $\gamma(\lambda)\ll 1 \, , \, \gamma(\lambda)=\hat \delta(\lambda)/(2\, \xi_1(\lambda))=O(q^2)$ when $\delta_2 \to 0$. 
\section{Considerations on the existence of additional fixed points for the Hamiltonian~\texorpdfstring{\eqref{eq:ham_det1}}{}} \label{sec:app3}
The determination of the fixed points of the Hamiltonian~\eqref{eq:ham_det1} is done by imposing that $\dot \phi=0$ and $\dot J=0$, \ie
\begin{equation}
    \begin{split}
        \cos \phi & = 0 \\
        \delta + 2 \alpha J \pm \frac{q}{2} \frac{1-2J}{\sqrt{J(1-J)}} & = 0 \, ,
    \end{split}
\end{equation}
where the first equation gives $\phi=\pi/2, 3\pi/2$ and the second one the following quartic equation
\begin{equation}
\begin{split}
    16 \, \alpha^2 J^4+16 \, \alpha(\delta -\alpha)J^3+&4(q^2+\delta^2-4\alpha \delta)J^2 + \\
    & -4(q^2+\delta^2) J +q^2 = 0 \, ,
\end{split}
\label{eq:fp}
\end{equation}
where it is immediate to observe that the coefficients of the terms $J^4$, $J$, and the constant term have fixed sign, which is positive, negative, and positive, reflectively. On the other hand, the coefficients of the terms $J^3$ and $J^2$ do not have a fixed sign, but when $\delta \ll 1$ the first is negative and the latter is positive.

For our purpose, we want to determine the conditions under which Eq.~\eqref{eq:fp} has two real solutions, which occurs when the discriminant $\Delta$ is negative. By direct computations, one obtains
\begin{equation}
    \begin{split}
        \Delta = &-65536 \, \alpha ^2 q^2 (\alpha +\delta )^2 [27 \alpha ^4 q^2+\alpha ^3 (8 \delta ^3+54 \delta  q^2)+\\
        +& \alpha ^2 (12 \delta ^4+39 \delta
   ^2 q^2)+6 \alpha  \delta  (\delta
   ^2+q^2)^2+(\delta ^2+q^2)^3] \\
  = & -65536 \, \alpha ^2 q^2 (\alpha +\delta )^2 f(\alpha)
    \end{split}
    \label{eq:discr}
\end{equation}
and this implies that $f(\alpha)$ should always be positive. Note that $f(\alpha)$ is also represented by a quartic polynomial and its sign can be studied by considering its discriminant, $\hat \Delta$, which reads
\begin{equation}
    \begin{split}
        \hat \Delta = &314928 \, q^4 (q^4-\delta ^4)^3 (\delta ^4+4 q^4)^2
    \end{split}
    \label{eq:discr1}
\end{equation}
and whose sign is easily determined
\begin{equation}
        \hat \Delta > 0 \qquad \text{if} \qquad -q < \delta < q \, .
    \label{eq:discr1sign}
\end{equation}

Therefore, considering also the following properties
\begin{equation}
\begin{split}
    f(0) & >0 \\ 
    f'(0) & > 0 \qquad \text{if} \qquad \delta > 0 
\end{split}
\end{equation}
one has that 
\begin{equation}
\begin{split}
    \text{if} & \quad \delta< -q  \;\; \text{or} \;\; \delta > q \quad \text{then} \quad \hat \Delta < 0 \\
    \text{hence} &\\
    & f(\alpha_i)=0 \quad i=1,2 \;\; \alpha_i \in \mathbb{R} \\
    \text{and} & \\
     & \delta < -q \quad 0< \alpha_1 < \alpha_2 \\ 
     & \delta > q \quad \alpha_1 < \alpha_2 < 0 \\ 
     \text{hence} &\\ 
     & f(\alpha)> 0 \quad \text{for} \quad \alpha<\alpha_1 \quad \text{or} \quad \alpha > \alpha_2 \, .
\end{split}
\end{equation}

Note that during the resonance-crossing process $\delta \to 0$ with constant $q$ and therefore $\hat \Delta$ will eventually change sign. 

Whenever $\hat \Delta > 0$ four or no real roots are possible and this depends on conditions on two additional quantities, namely 
\begin{equation}
\begin{split}
    \text{if} & \quad -q < \delta < q \quad \text{then} \quad \hat \Delta > 0 \\
    \text{hence if} &\\
    & \phantom{-}16 \, \delta ^6+27 \, \delta ^2 q^4 > 0 \\ 
    \text{and} & \\ 
    & -64 \, \delta ^{12}+8019 \, \delta ^4 q^8-216 \, \delta ^8 q^4+6561 \, q^{12} < 0\\ 
    & \text{there are four distinct roots} \\
    \text{else, if} & \\
    & \phantom{-}16 \, \delta ^6+27 \, \delta ^2 q^4 < 0 \\ 
    \text{or} & \\ 
    & -64 \, \delta ^{12}+8019 \, \delta ^4 q^8-216 \, \delta ^8 q^4+6561 \, q^{12} > 0\\ 
    & \text{there are no real roots}
\end{split}
\end{equation}
from which the conclusions on the number of real solutions of the equation $f(\alpha)=0$ depends only upon the study of the sign of the polynomial
\begin{equation}
   g(\delta)= -64 \, \delta ^{12}-216 \, \delta ^8 q^4+8019 \, \delta ^4 q^8+6561 \, q^{12}
\end{equation}
that is even, \ie $g(\delta)=g(-\delta)$ and $g(\delta) \to -\infty $ for $\delta \to \pm \infty$. $g(\delta)$ can be considered as a cubic polynomial in the variable $\tilde \delta=\delta^4$ and one can verify that the discriminant is always positive, thus ensuring that there are three real and distinct roots of the equation $g(\tilde \delta)=0$. Moreover, their product is positive, thus imposing that they are all positive or one positive and two negative. It turns out that indeed only one is positive and the solution of the initial equation is given only by
\begin{equation}
    \delta^4 = \mu q^4 \quad \text{\ie} \quad \delta = \pm \mu^{1/4} |q|  \, , \quad \mu \approx 17.9085 \times \frac{9}{16} \, .
\end{equation}
Therefore
\begin{equation}
    g(\delta)> 0 \quad \text{if} \quad -\mu^{1/4} q < \delta < \mu^{1/4}q
\end{equation}
and a fortiori it is positive in the interval $[-q,q]$, which shows that on that interval $f(\alpha)=0$ has four real and distinct roots. It is easy to verify that an even number of roots can be negative and such a number does not depend on the sign of $\delta$. 

On the other hand, looking at the extrema of $f(\alpha)$, one finds three real and distinct extrema and the product of the values $\alpha_{i, \rm ext}, i=1,2,3$ for which the extremum is reached satisfies $\text{sgn}(\alpha_{1,\rm ext}\alpha_{2,\rm ext}\alpha_{3,\rm ext})=-\text{sgn}(\delta)$, which shows that for $\delta>0$ an odd number of extrema is negative, whereas for $\delta<0$ an even number is negative. 

It is then clear that as the sign of the solutions of $f(\alpha)=0$ does not depend on the sign of $\delta$ while the sign of $\alpha_{i, \rm ext}$ does, only two solutions $\alpha_i$ are positive and the number maxima with negative position varies from two for $\delta < 0$ to one for $\delta>0$. In summary, the following holds
\begin{equation}
\begin{split}
    \text{if} & \quad -q < \delta < q \quad \text{then} \quad \hat \Delta > 0 \\
    \text{hence} &\\
    & f(\alpha_i)=0 \quad i=1,2,3,4 \;\; \alpha_i \in \mathbb{R} \\
    \text{and} & \\
     &  \alpha_1 < \alpha_2 < 0 < \alpha_3 < \alpha_4 \\ 
     \text{hence} &\\ 
     & f(\alpha)> 0 \;\; \text{if} \;\; \alpha<\alpha_1 \;\; \text{or} \;\; \alpha_2 < \alpha < \alpha_3 \;\; \text{or} \;\; \alpha > \alpha_4
\end{split}
\end{equation}
and this shows that it is indeed possible to have only two fixed points in phase space with a non-zero detuning with amplitude.
\section{Detail of the computation of the map used in the numerical simulations}\label{sec:app4}
The matrix $\bm{M}_\text{FODO}$ can be transformed in Jordan form via the transformation $\bm{T}$, so that $\bm{T}^{-1}\bm{M}_\text{FODO}\bm{T} = \bm{R}(\omega_x,\,\omega_y)$, where $\bm{R}(\omega_x,\,\omega_y)$ is the 4D rotation matrix for the frequencies $\omega_x$ and $\omega_y$, \ie
\begin{equation}
\bm{R}(\omega_x,\,\omega_y)= \bm{R}(\omega_x) \otimes \bm{R}(\omega_y) = \begin{pmatrix} \bm{R}(\omega_x) & \bm{0} \\ \bm{0} & \bm{R}(\omega_y)\end{pmatrix} \, ,
\end{equation}
where $\bm{R}(\omega_z)$ is a standard 2D rotation matrix. This transformation induces a  new set of co-ordinates $\bm{X}=(X,\,X',\,Y,\,Y')$ defined as $\bm{X}=\bm{T}^{-1}\bm{x}$ where $\bm{T}$ is well-known and reads
\begin{equation}
\bm{T} = \begin{pmatrix} \bm{T}_x & \bm{0} \\ \bm{0} & \bm{T}_y\end{pmatrix} \qquad\text{where } \bm{T}_z = \begin{pmatrix} \sqrt{\beta_z} & 0 \\ -\alpha_z/\sqrt{\beta_z} & 1/\sqrt{\beta_z} \end{pmatrix} \, .
\end{equation}

The map then reads
\begin{equation}
    \begin{split}
        \bm{X}_{n+1} &= \qty(\bm{T}^{-1}\bm{x}_{n+1})\\
        &= \qty(\bm{T}^{-1} \bm{M}_\text{FODO}\bm{T})   \qty(\bm{T}^{-1} \bm{M}_\text{Skew}\bm{T})\qty(\bm{T}^{-1}\bm{x}_{n}) \\
        &=\bm{R}(\omega_x,\omega_y)\qty(\bm{T}^{-1}\bm{M}_\text{Skew}\bm{T})\bm{X}_n \, ,
    \end{split}
\end{equation} 
where 
\begin{equation}
\hat{\bm{M}}_\text{Skew}=\bm{T}^{-1}\bm{M}_\text{Skew}\bm{T} = \begin{pmatrix} \bm{1} & \bm{Q} \\ \bm{Q} & \bm{1}\end{pmatrix}\end{equation}
and
\begin{equation}
\bm{Q} = \begin{pmatrix} 0 & 0 \\ q & 0 \end{pmatrix},\qquad q = \sqrt{\beta_x\beta_y} \hat q
\end{equation}
so that
\begin{equation}
\bm{X}_{n+1} = \bm{R}(\omega_x,\,\omega_y)\hat{\bm{M}}_\text{Skew}= \begin{pmatrix} \bm{R}(\omega_x) & \bm{Q}_x \\ \bm{Q}_y & \bm{R}(\omega_y)\end{pmatrix}\bm{X}_n
\label{eq:mapRQ}
\end{equation}
having defined
\begin{equation}
\bm{Q}_z = q \begin{pmatrix} \sin\omega_z & 0 \\ \cos\omega_z & 0 \end{pmatrix} \, .
\end{equation}

\newpage
\bibliographystyle{unsrt}
\bibliography{mybibliography}
\end{document}